\documentclass[a4paper]{article}
\usepackage[a4paper, total={6in, 8in}]{geometry}
\usepackage{setspace}
\usepackage{lineno}

\usepackage{standalone}
\usepackage{authblk}

\usepackage{amsmath}
\usepackage{amssymb}
\usepackage{url}
\usepackage{booktabs}
\usepackage{caption}
\usepackage{multirow}

\usepackage{graphicx}
\graphicspath{{./figs/}}
\usepackage[caption=false]{subfig}


\makeatother

\usepackage{color,soul}

\usepackage{comment}
\usepackage{algorithm}
\usepackage{algpseudocode}

\algnewcommand\algorithmicforeach{\textbf{for each}}
\algdef{S}[FOR]{ForEach}[1]{\algorithmicforeach\ #1\ \algorithmicdo}

\usepackage{algpseudocode}
\algrenewcommand\textproc{}

\let\oldReturn\Return
\renewcommand{\Return}{\State\oldReturn}

\DeclareMathOperator*{\argmax}{arg\,max}

\usepackage{silence}
\begin{document}

\begin{flushleft}
{\huge The emergence of division of labor through decentralized social sanctioning}  \bigskip

{\normalsize Anil Yaman\textsuperscript{1\ddag*}, Joel Z. Leibo\textsuperscript{2\ddag}, Giovanni Iacca\textsuperscript{3}, Sang Wan Lee\textsuperscript{4}}
\bigskip

{\small \textbf{1} Vrije Universiteit Amsterdam, Amsterdam, The Netherlands\\
\textbf{2} DeepMind, London, UK\\
\textbf{3} University of Trento, Trento, Italy\\
\textbf{4} Korea Advanced Institute of Science and Technology, Daejeon, Republic of Korea
}
\bigskip

{\small \ddag These authors also contributed equally to this work.\\}
{\small *Corresponding author (AY): a.yaman@vu.nl}

\end{flushleft}


\section*{Abstract}
Human ecological success relies on our characteristic ability to flexibly self-organize into cooperative social groups, the most successful of which employ substantial specialization and division of labor. Unlike most other animals, humans learn by trial and error during their lives what role to take on. However, when some critical roles are more attractive than others, and individuals are self-interested, then there is a social dilemma: each individual would prefer others take on the critical but unremunerative roles so they may remain free to take one that pays better. But disaster occurs if all act thusly and a critical role goes unfilled. In such situations learning an optimum role distribution may not be possible. Consequently, a fundamental question is: how can division of labor emerge in groups of self-interested lifetime-learning individuals? Here we show that by introducing a model of social norms, which we regard as emergent patterns of decentralized social sanctioning, it becomes possible for groups of self-interested individuals to learn a productive division of labor involving all critical roles. Such social norms work by redistributing rewards within the population to disincentivize antisocial roles while incentivizing prosocial roles that do not intrinsically pay as well as others.



\section{Introduction}

Human societies depend on division of labor. However, an individual's role is not specified in their genes. Rather, human roles are learned during individual lifetimes. This is one of the reasons why human groups can achieve collective welfare more quickly than what would be possible with purely genetic evolution. However, many computational models of lifetime learning formulate it as a process of maximization of individual payoffs~\cite{lee2019decision,lee2012neural,dayan2008reinforcement,sutton2018reinforcement}. Such a formulation cannot, on its own, account for the learning of division of labor. When individuals are driven by self interest they cannot learn to perform roles that do not pay as well as other roles, yet such roles are often necessary for the group to function and achieve a high overall welfare. Consequently, a fundamental question arises: how can division of labor emerge in groups of self-interested lifetime-learning individuals?

Here, we hypothesize that social norms, which we take to be patterns of social sanctioning, are sufficient to incentivize individuals in groups to select prosocial role choices, thereby enabling group-level division of labor to emerge from self-interested lifetime-learning. To test this hypothesis, we propose a model where lifetime-learning is shaped by social sanctioning. The specific social sanctioning mechanism presented here was inspired by studies that investigated the role of reward and punishment in laboratory-based social dilemmas such as the public goods game~\cite{fehr2000cooperation,fehr2002altruistic,perry2018collective,albrecht2018peer,boyd1992punishment}.

In these experiments, participants start with an initial endowment of tokens and can contribute to a public good by investing tokens in it. It is best for all if all individuals invest all their tokens, yet as individuals, all face an incentive to free-ride: holding back their own investment while still benefiting from the contributions of others. For instance, Fehr and G\"achter~\cite{fehr2002altruistic} designed experiments consisting of two stages. First, participants chose how much to contribute to the public good. Then in a second stage, they received information about the choices of others, and then decided on that basis whether or not to pay a cost to punish others for their behavior. They found that participants were willing to engage in altruistic punishment: they would pay to punish free riders. Furthermore, the possibility of altruistic punishment in the case of repeated interactions led to an increase in the average cooperation level of the participants in the group.

\begin{figure*}[!ht]
 \centering
 \includegraphics[width=0.95\textwidth]{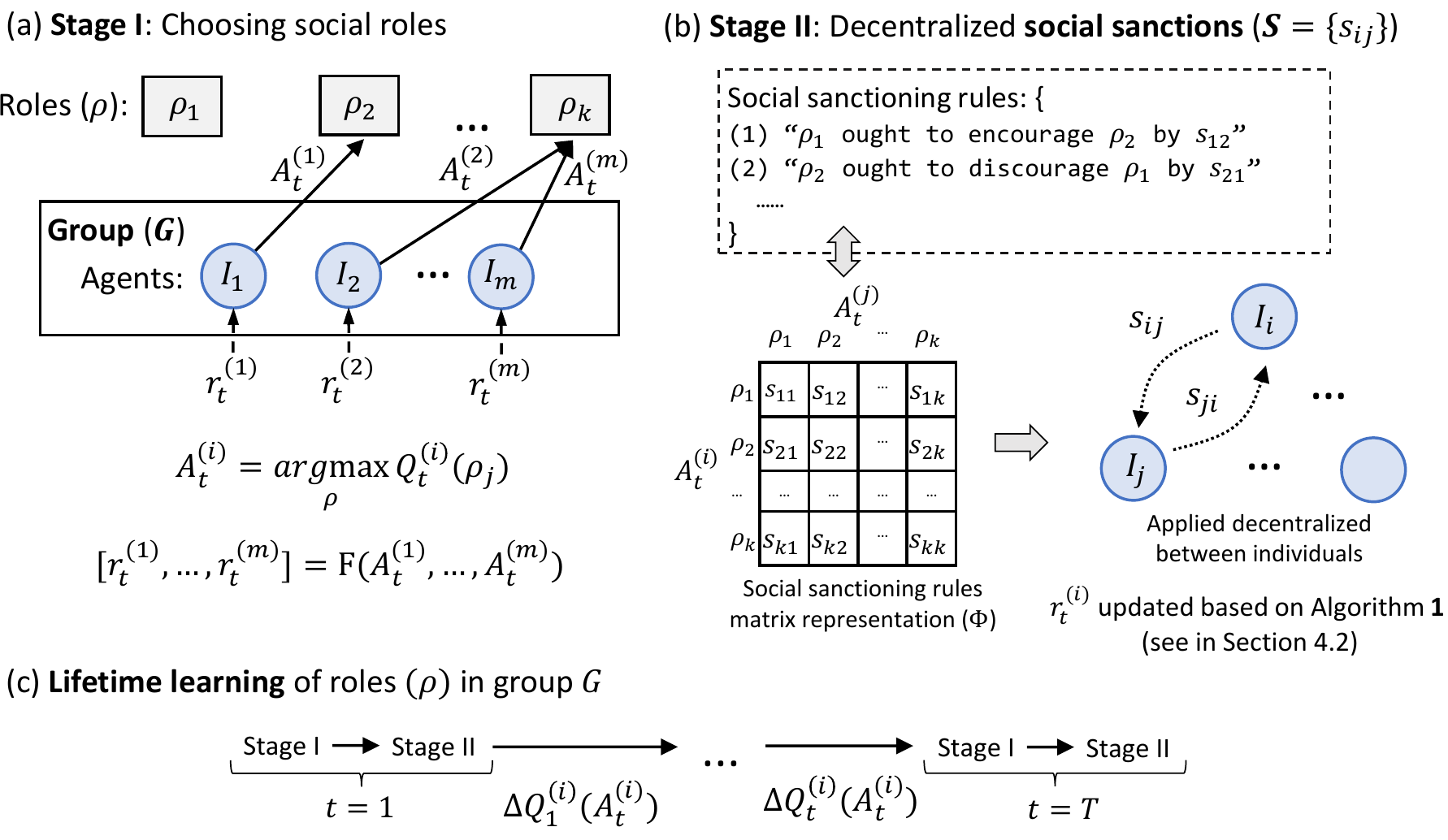}
\caption{The decision process of the individuals consists of two stages: in (a) \textbf{Stage I}, individuals make a role selection based on the values of the roles ($Q^{(i)}_t(\rho)$) (see Equation (1)). The rewards received depend on a function ($F$) of the roles selected by all individuals in the group (see Equation (3)). In (b) \textbf{Stage II}, social sanctioning rule examples are shown in forms of statements given in the box outlined with dashed lines. Considering $k$ number of roles, it is possible to generate $k\times k$ number of rules. For computational efficiency, we represent these rules as a matrix where each cell encodes the amount of encouragement/discouragement is imposed by corresponding roles indicated in the row and column headers. Social sanctions can take forms negative (in case of reward), positive (in case of punishment), or zero (if no social sanction is imposed). Social sanctions are imposed decentralized fashion to shape the rewards of others to encourage the division of labor. The rewards are updated based on Algorithm 1 given in Section 4.2. Such a pattern of decentralized social sanctioning constitutes a social norm ($\boldsymbol{S}$). The learning process consists of two levels: lifetime-learning and social norm evolution. (c) In lifetime-learning, groups of individuals perform Stage I and Stage II iteratively to learn their roles. Through social norm evolution, social sanctioning matrix values are optimized to produce optimum lifetime-learning of the social roles.}
 \label{fig:socialSanctions}
\end{figure*}

Figure~\ref{fig:socialSanctions} illustrates the learning process used in our model. In \textbf{Stage I} (see Figure~\ref{fig:socialSanctions}a), a group of self-interested lifetime-learners learn their roles. This process is modeled as a $K$-armed bandit problem where individuals learn the role (arm) that provides the highest payoff (optimum) from a finite set of roles. Each individual's lifetime-learning of their role is modeled by an $\epsilon$-greedy algorithm: individuals select the role they judge to provide the maximum average payoff as estimated empirically by the payoffs they receive and explore other roles with a small probability $\epsilon$~\cite{sutton2018reinforcement}. \textbf{Stage II} (see Figure~\ref{fig:socialSanctions}b) introduces the social sanctioning stage where individuals can monitor others, and impose sanctions based on their role choices. Subsequently, the payoffs received in Stage I are updated by the social sanctions imposed in Stage II.

In our model, social sanctioning occurs because individuals endorse normative rules of the form ``X ought to encourage/discourage Y". For example ``Farmers ought to encourage Hunters''. Encouragement is not possible when individuals have no resources, so individuals with no recent rewards do not encourage---regardless of what their society deems normative for their role. In this model costly norms cannot be implemented by impoverished individuals, however the rules themselves are not forgotten and may reactivate once individuals with the roles they name become wealthier. Likewise, systems of norms retain rules for social roles that may not exist. Rules for how farmers ought to relate to soldiers can persist in the cultural memory even if soldiers do not currently exist.

We model social norms as patterns of social sanctioning adopted by all individuals in a group. That is, every individual in the group applies the same sanctioning scheme. This is consistent with conceptions of what it is for a collective behavior pattern to be a social norm that rely on most agents in the group to conform~\cite{bicchieri2018social}. This kind of sanctioning mechanism is said to be decentralized since it does not require a centralized enforcement mechanism like a government-run police force. Rather, the individuals in the group perform all the sanctioning actions themselves. However, their participation is not voluntary. They do not decide on their own whether to sanction or not. At each point in time, whether they sanction another individual, and if so, how much, is entirely determined by the group's social sanctioning matrix, i.e., the social norm itself. This may be justified by assuming the existence of a metanorm that demands individuals sanction in accord with their group's overall pattern, an assumption also shared by other models where norms are seen as public goods and metanorms are consequently necessary to evade the second-order free-rider problem~\cite{axelrod1986evolutionary, heckathorn1989collective}. The existence of the requisite metanorm to stabilize patterns of decentralized social sanctioning is supported by laboratory experiments~\cite{yamagishi1986provision, horne2002sanctioning, vaish2016preschoolers, weber2018dispositional} and ethnographic evidence~\cite{mathew2017second, eriksson2021perceptions}. In our model, one implication of involuntary social sanctioning is that, given the right norm, it is possible to shape learning in any direction. This works for the same reason that reward shaping techniques are effective both in behaviorist psychology~\cite{peterson2004day} and in algorithmic reinforcement learning~\cite{erez2008does}. Here groups can induce individuals by sanctioning to select any behavior. This is consistent with other models in the evolution of cooperation literature where sanctioning may stabilize behaviors regardless of whether or not they are adaptive~\cite{boyd1992punishment}.

As illustrated in Figure~\ref{fig:socialSanctions}b, social sanctions are determined by rules which are  computationally encoded in the form of a real-valued matrix with shape $k \times k$ (where $k$ is the number of roles). Entries in the matrix correspond to normative rules. Each entry defines how much reward or punishment an individual with role $i$ ought to impose on an individual of role $j$. Under our terminology, each specific such matrix corresponds to a distinct social norm. Sanctioning in this model is role-wise. The amount of sanctioning applied by one individual to another at time $t$ is a function of both of their roles at that time. There is a rich and cross-disciplinary tradition that centers around theories of social structure which features this kind of intimate connection between ``role psychology'' and normative behavior. Intuitively, norms are deeply entangled with social roles. For instance, it would be inappropriate (i.e.~sanctionable) for a student to behave like a teacher or for a judge to behave like a legislator~\cite{sunstein1996social, bicchieri2006grammar, eickers2022coordinating}. The errant student would likely face discipline from an academic administrator---another role, and one for which such sanctioning is part of its job description. Notice also that role-wise sanctioning---but not individual-targeted sanctioning---allows the implementation of \emph{impartial} norms, like some human norms are~\cite{hadfield2014microfoundations, railton2017moral}. It need not merely produce the self-serving resentment-driven sanctioning motivations that are sometimes observed in non-human primates~\cite{brosnan2003monkeys}.

Depending on the environment, a social norm may either facilitate high social welfare or court disaster. For instance, a norm that encourages many farmers to become soldiers could be important for survival in a hostile environment where soldiers are truly needed for defense. However, in a more benign environment, with no external threats for soldiers to defend against, then allocating too many individuals to the soldier role would be suboptimal, and it would be better to have more farmers instead. Typically the environment is a major factor in which specific norms are effective.

One implication is that norm evolution may grow more and more precisely optimized to local conditions over time. It is generally possible to achieve a better and better fit to the precise environmental conditions a community finds itself in by evolving more and more complex norms. However, we think there are at least two strong brakes on this process that prevent the evolution of greater and greater norm complexity. First, we must consider the fact that for cultural evolution to happen, norms must somehow be transmitted between generations. These conditions are thought to give rise to a simplicity bias (complexity regularization~\cite{hastie2009elements}) in the context of direct and indirect reciprocation rules for cooperation~\cite{rubinstein1986finite,santos2021complexity}, and language learning~\cite{tamariz2016cultural}. We propose that the same mechanisms also constrain norm-acquisition.  So if norm learners (children or immigrants) find it easier to acquire simpler norms, then there will be an overall bias toward simplicity throughout the evolutionary process. There is additionally one other kind of brake on complexity growth: very high norm complexity may be problematic in itself if we assume that it implies a growing overhead cost. Tainter (1988) argued that the tendency for states to accumulate complexity, with large attendant costs, has been a major causal force in the collapse of ancient empires~\cite{tainter1988collapse}. If true, this logic would also seem to put a brake on the benefits from increasing norm complexity.

In our model, social norms are optimized through a cultural evolution process (given in Algorithm~\ref{alg:normEvolution}). We interpret its mechanism of norm change as cultural group selection~\cite{waring2018evidence, henrich2004cultural, richerson2016cultural}. As such, our model depends critically on all the assumptions necessary for the strength of group selection to outweigh that of individual selection e.g.~sufficient separation between groups~\cite{henrich2004cultural}. Note though that in the present work we do not explicitly model other groups beyond the focal group. Formally, our model considers only a single representative group where the effect of evaluating and evolving social norms can be computed iteratively and independently. Each social norm evolution experiment starts from a randomly initialized norm, obtained by sampling the values of the sanctioning matrix from a uniform distribution over a certain range. In each iteration, a new variant of the social norm is generated by perturbing its constituent rules. The newly generated norm is then evaluated, and its success measured by average group payoff achieved on a task demanding the learning of a division-of-labor arrangement. In addition, we introduce complexity regularization~\cite{hastie2009elements,louizos2017learning} that impose a penalty based on the complexity of the norms. The evolutionary process of the norms starts from an initial condition that does not incorporate any prior knowledge of the task at hand. If the norm achieves a higher average group payoff then it replaces the status quo.

Overall, we find that the social norms that emerge from evolution involve redistribution mechanisms where individuals periodically pay others to incentivize them to perform beneficial roles for the group that they would not otherwise select. Consequently, these mechanisms allow groups of self-interested individuals to discover effective division-of-labor arrangements through lifetime-learning. Specifically, we show that the proposed method of simulating social norm evolution leads to higher collective payoff than groups of self-interested or altruist individuals. Moreover, complexity regularization promotes convergent evolution to simpler social norms in independent evolutionary processes.

\section{The Model}

\subsection{Lifetime-learning of individuals' roles}

In our model, groups consist of individuals. Individuals select and re-select roles for themselves throughout their lifetime. Individuals learn from this experience which of their role choices are the most rewarding and usually select the role they expect will provide the most reward. That is, individuals face a $K$-armed bandit problem $\rho = \{\rho_1, \rho_2,\hdots, \rho_K\}$ where $K$ is the number of roles. They select a new role on each step, so it is best to think of the simulation steps as corresponding to some substantial period of time like a week or a month.

The lifetime-learning process is illustrated in Figure~\ref{fig:socialSanctions}a. Individuals change their behavior over time via value-based reinforcement learning~\cite{sutton2018reinforcement}. On each step $t$, each individual $i$ selects its role $A_t^{(i)}$ using its estimated value $Q^{(i)}_t(\rho_j)$ with
\begin{equation}
\label{eq:roleSelection}
A^{(i)}_{t} = \argmax_{\rho} Q^{(i)}_{t}(\rho_j) 
\end{equation}
and updates value estimates using the reward received on the previous step, as shown below:
\begin{equation}
\label{eq:ILrewardUpdate}
Q^{(i)}_{t+1}(A^{(i)}_{t}) = Q^{(i)}_{t}(A^{(i)}_{t}) + \alpha \left[r^{(i)}_t - Q^{(i)}_{t}(A^{(i)}_{t}) \right]
\end{equation}

where $0< \alpha\leq 1$ is the learning rate parameter and $r^{(i)}_t$ is the reward received after selecting $A^{(i)}_{t}$. Individuals aim to maximize their rewards and thus usually select the role they estimate to have the highest value, occasionally also exploring other roles with a small probability $\epsilon$. We measure an individual's performance after $T$ iterations. 

In this model, individuals learn to select roles in order to maximize their personal reward. However, this could lead them to select roles containing selfish behaviors that gain personal reward at the expense of the wider group. Individualistic reinforcement learning has the effect of discouraging agents from ``taking one for the team'', resulting in lower joint performance in environments that require such cooperation.

\subsection{Incentivizing the lifetime-learning of division of labor via social norms}

We consider a reward function $F$ in the form shown in Equation~\eqref{eq:rewardFunction} where the reward $r^{(i)}_t$ received by individuals $i$ at time step $t$ depends on the roles selected by all individuals in the group, not just its own choice. Individuals' rewards may be interdependent. In situations of interdependence, self-interested optimization of personal reward often does not converge to a socially optimum role distribution since some roles that are critical for optimal group welfare are not as individually rewarding as other roles, so no individuals learn to select them.

\begin{equation}
\label{eq:rewardFunction}
\left[r^{(1)}_t, \hdots, r^{(m)}_t \right] = F(A^{(1)}_t, \hdots, A^{(m)}_t)
\end{equation}

Social norms in our model are regarded as decentralized patterns of social sanctioning. In our model, sanctions are rewards and punishments imposed by one individual on another individual. Amounts of sanctioning are a function of the roles of the sanctioning and the sanctioned player. The social sanctioning stage (Stage II as shown Figure~\ref{fig:socialSanctions}b) is applied every time step after Stage I (shown in Figure~\ref{fig:socialSanctions}a). Here, the individuals in the group monitor other individuals (based on a certain neighborhood function that defines the connectivity of their social network) and impose sanctions. The amount of the sanction provided by an individual taking role $k$ to an individual taking role $\ell$ is $s_{k, \ell}$ (see Figure~\ref{fig:socialSanctions}b). The rewards received after the role selection (Stage I) are then updated based on the social sanctioning scheme proposed in Stage II. As shown in Figure~\ref{fig:socialSanctions}c, Stage I and II are performed consecutively for $T$ iterations.

\subsection{Evolution of social norms for incentivizing division of labor}\label{sec:culturalEvoSocialNorms}

We formalize the process of social norm evolution with an optimization algorithm that models cultural evolution from the cultural group selection point of view (see Algorithm~\ref{alg:normEvolution}). The goal of the algorithm is to find a particular set of social sanctioning rules, represented as matrix $\boldsymbol{S^*}$, that can maximize: $$\boldsymbol{S^*} = \argmax_{\boldsymbol{S}} (Ft({\boldsymbol{S}}))\text{,}$$ where $Ft(\boldsymbol{S})$ is the \textit{fitness} of a norm $\boldsymbol{S}$ composed of: $$Ft(\boldsymbol{S}) = R - \lambda \lVert \boldsymbol{S} \rVert_0\text{,}$$ where $R$ is the average group payoff received by the individuals as a result of the role distributions they learn, $- \lambda \lVert \boldsymbol{S} \rVert_0$ is the \textit{complexity regularization} (also known as $L_0$ regularization~\cite{hastie2009elements,louizos2017learning}), where $\lVert \boldsymbol{S} \rVert_0$ denotes the number of non-zero values in social sanctioning matrix and $\lambda$ is the parameter to adjust the weight for the importance given to the complexity regularization during the optimization process. Given two norms that can achieve similar average reward, the one that consists of a smaller number of rules would be simpler in terms of their understandability and computational complexity. When $\lambda>0$, complexity influences the overall fitness of social norms and favors those that are simpler. Higher $\lambda$ values are expected to increase the selection pressure for favoring simpler social norms. The sensitivity analysis on $\lambda$ for the evolutionary processes modeled in this paper is presented in Supplementary Material.

Other forms of complexity measures can be considered. For instance, the sum of the absolute values of the social sanctioning values (i.e. $L_1$ norm) could be a good measure for minimizing the total amounts of sanctioning values. On the other hand, this may not reduce the number of rules that constitute the social norms since there is still a possibility of the emergence of large number of rules with very small sanctioning values. Other option may be the number of types of roles involved in the rules. In this case, there is also a possibility of the emergence of social norms that consist of large number rules but contain a small number of role types. Therefore, $L_0$ norm is a better fitting measure among these alternatives for modelling the complexity of the social norms that can facilitate to the emergence of simpler norms in terms of their size.

Initially, the group is assigned a social sanctioning matrix where the amount of sanctions are randomly sampled within a certain range (i.e., uniformly in $[-6,6]$). It is assumed that all individuals in the group use the same social sanctioning matrix. Then, the group can sample a new social norm by applying four possible operators to the existing one: (1) Gaussian perturbation: the values of non-zero cells perturbed by $\mathcal{N}(0,\sigma)$, (2) addition of a new rule: a randomly selected zero-valued cell initialized from $\mathcal{N}(0,1)$ and (3) deletion of an existing rule: the value of a randomly selected non-zero cell is replaced by 0. Note that addition and deletion operators cause changes in the number of sanctioning rules whereas, Gaussian perturbation causes only a change in the sanctioning values of existing rules. In addition, we keep track of the usage frequency of each rule (i.e. how many times each rule is implemented by the individuals in a group), and if there are rules that are not used, we remove them by assigning their amount to 0 in the social sanctioning matrix.

We introduce four parameters to adjust the amount of change these operators can lead to. These parameters: \textit{mutation probability (mp)} and \textit{mutation rate (mr)} for Gaussian perturbation, \textit{rule addition probability (rap)} for rule addition, and \textit{rule deletion probability (rdp)} for rule deletion. Gaussian perturbation is applied with the probability of $mp$ and rate of $mr$ where $\sigma = mr$, and rule addition and rule deletion operators are applied based on the probabilities of $rap$ and $rdp$ respectively. Sensitivity analysis for these parameters is provided in Supplementary Material.

After the perturbation step, if the perturbed social sanctioning rules $\boldsymbol{S^{\prime}}$ provides better performance (as measured by $Ft$), it is selected by the group. It then replaces the social sanctioning rules currently in use. After a certain number of iterations of this process we expect to find social sanctioning rules that are better at incentivizing division of labor.

The average reward of the social sanctioning rules $R$ is found as follows: 
\begin{equation}
 \label{eq:lifetimeReward}
 R = \text{EVAL}(\boldsymbol{S}) = \frac{1}{M} \sum^M_{m=1} \sum^T_{t=\zeta} r^{(m)}_{t} 
\end{equation}
where $\text{EVAL}()$ is the evaluation function of social sanction $\boldsymbol{S}$, $M$ is the number of individuals in the group, and $\zeta$ is cutoff for computing the rewards. The problems the groups tackle require the individuals to learn various roles so that $R$ can be maximized. The evaluation function is concerned with the lifetime-learning process of individuals and repeated for $T$ steps as illustrated in Figure~\ref{fig:socialSanctions}. The lifetime-learning process is stochastic since it depends on the role choices (i.e., random exploration of roles and the order of sanctions). Therefore, to rule out randomness, the evaluation is performed multiple times and the average is accepted as the final result.

\subsection{Spatial games}

\begin{figure*}[!ht]
 \centering
 \includegraphics[width=0.98\textwidth]{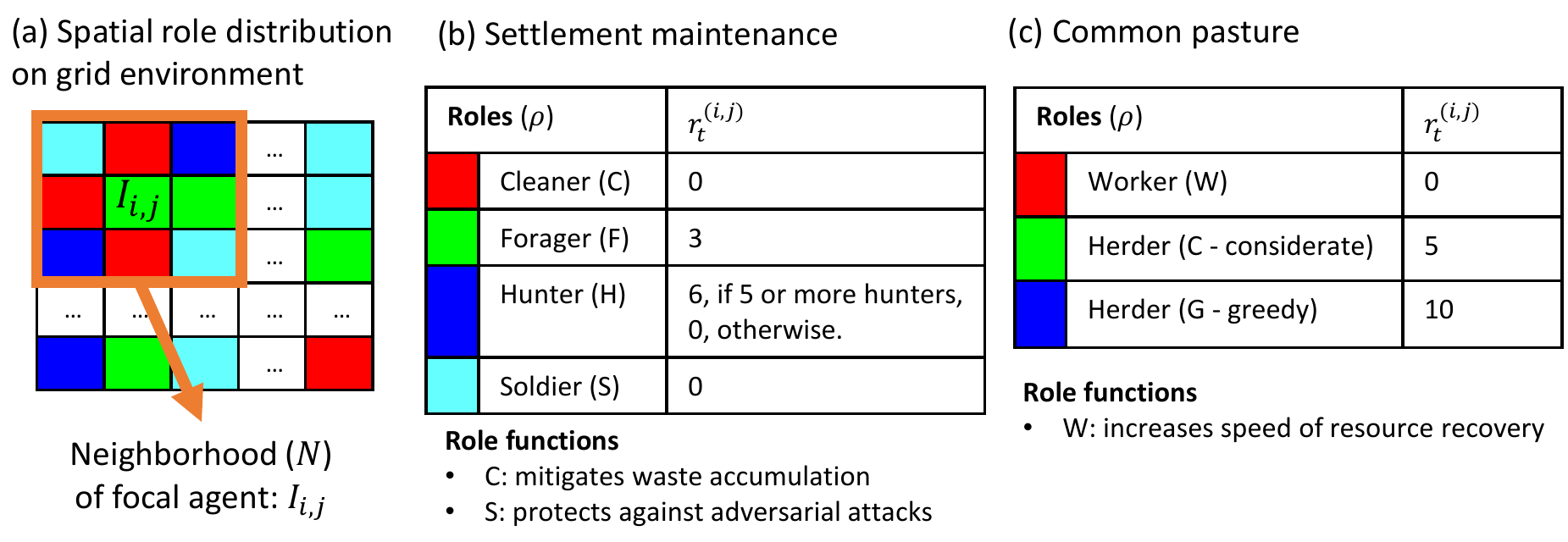}
 \caption{(a) Types of roles, color coded for visualization, can be performed by the individuals that are spatially distributed on a 2-dimensional toroidal grid environment and can impose social sanctions to the individuals within their neighborhood. The goal is to learn a role distribution (a role in each location) that can maximize the average group payoff. The payoffs of the roles are shown in Tables (b) and (c) in settlement maintenance and common pasture games respectively. The settlement maintenance has four and common pasture has three role choices.}
 \label{fig:payoffMatrices}
\end{figure*}

We investigate the evolution of cooperative norms on two spatially structured games called \textit{settlement maintenance} and \textit{common pasture}. In both games, each individual occupies a position in space and directly interacts only with their neighbors (see Figure~\ref{fig:payoffMatrices}a). Each individual selects a role to perform, and receives a payoff as a result. Some of the roles do not provide individuals who select them with any reward at all, yet it still may be necessary that some agent perform them for the group as a whole to be successful.

We indicate the locations (cells) that individuals are assigned by $(i,j)$ coordinates. They are constrained to interact with their closest eight neighbors (i.e., their Moore neighborhood). Furthermore, the environment is toroidal so that it allows the individuals located at the first and last columns/rows to interact with the individuals located at the last and first columns/rows.

The settlement maintenance game was inspired by research on ant colonies but interpreted here as a model for small-scale human societies~\cite{gordon1996organization,lichocki2012neural}. Payoffs associated to each role are shown in Figure~\ref{fig:payoffMatrices}b. In this game, individuals can obtain payoffs by perform foraging and hunting roles in form of food. Hunting provides a higher payoff but requires coordination with a sufficient number of other hunters. In addition, there are cleaner and soldier roles that do not directly provide any payoffs but are still required for maintaining and protecting the settlement respectively. In each time step, there is a constant rate of waste accumulation in each cell of the environment which negatively affects the rewards received by the individuals. However, if an individual performs the cleaner role in the neighborhood then the adverse effect of waste accumulation mitigated. Furthermore, there is an adversarial attack probability for each cell that causes the reward of an individual to be stolen (reduced to 0). However, when individuals take on the soldier role they then provide protection against adversarial attack within their neighborhood.

The payoffs of the roles in the common pasture game are provided in Figure~\ref{fig:payoffMatrices}c. This game models situations of resource appropriation where a bad outcome known as the tragedy of the commons may occur~\cite{Hardin1243,ostrom1990governing}. Here, the tragedy is when resources in the environment become depleted as a result of individuals pursuing their own selfish interests without regard to mounting social cost. To avoid the tragedy of the commons, a critical fraction of individuals must learn to act sustainably, restraining their use of resources. To simulate this behavior, we model a pasture environment with a certain amount of starting resources in each location (e.g. grass for the herders' herd). These resources can be appropriated by the herders either greedily (G) or considerately (C). The resources regenerate over time unless fully depleted. In addition, we introduce another role: worker (W) which can increase the regeneration speed of the resources (e.g. speed of grass growth). However, if the resource on a cell is fully depleted, it cannot be recovered within the present lifetime. Therefore, the individuals cannot receive any reward when the resources are depleted. The environment resets in terms of its resources and recovery ability between generations.

This model of social sanctioning is broadly compatible with other redistributive (zero-sum) sanctioning mechanisms in the reinforcement learning literature~\cite{lupu2020gifting, wang2021emergent, willis2023resolving}. It differs from the purely destructive negative-sum social sanctioning schemes considered by~\cite{vinitsky2023learning, koster2022spurious}. In the present work, the payoffs received in Stage I stand in for physical objects such as resources like the food received as a result of foraging behavior. This kind of sanctioning is called zero-sum because it respects a conservation law: resources are neither created nor destroyed by sanctioning. Consequently, the function of sanctioning can be considered as a decentralized resource redistribution scheme. Losing resources is punishing while gaining resources is rewarding. One may ``gift'' resources to another, reducing your own reward to increase theirs (positively sanctioning them). Symmetrically, one may take resources away from another agent, gaining them yourself (negatively sanctioning them). Importantly, once reward has been reduced to zero it cannot be reduced any further since one cannot take away resources that do not exist. One implication is that individuals can only positively sanction others when they have enough resources to do so. The sanctioning process is shown in detail in Algorithm~\ref{alg:rewardUpdateSocialSanctions}.

\subsubsection{Settlement maintenance}

In the settlement maintenance game, individuals can select one of the roles in each time step $t$ as: $A^{(i)}_t\in \{$\textit{cleaner, forager, hunter, soldier}$\}$. We assume there is a level of waste accumulation $d^{(i,j)}_{t+1} = \min(wa, d^{(i,j)}_t + \tau)$ in each cell location $(i,j)$ with a constant rate of $\tau$, reaching a maximum level. The ideal living conditions of the individuals are affected by the waste accumulation. Thus, their payoffs are updated based on the waste accumulation as follows:
\begin{equation}
 r^{ (i,j)}_t = \max (0, r^{\prime (i,j)}_t - d^{(i,j)}_t).
\end{equation}
where $r^{\prime (i,j)}_t$ is the payoff received by the individuals. On the other hand, the negative effect of waste accumulation is mitigated by the cleaner individuals by setting the rate of waste accumulation to 0 within the neighborhood of the cleaner individuals.

In addition, we assume that the environment is subject to random adversarial attacks that reduce the payoffs received by the affected individuals. The frequency of these attacks is controlled by the attack probability $P_{i,j}$ that is assumed to be the same for all cells. If there is an attack on a cell location, the payoff of the individual is reduced to 0. On the other hand, if there is a soldier in the neighborhood, the attacks are defended and payoffs are protected.

\subsubsection{Common pasture}

In common pasture, individuals can select one of the roles as : $A^{(i)}_t\in \{$\textit{considerate, greedy, worker}$\}$. The amount of resources in each cell represented as $D^{(i,j)}$. In the beginning of the game, certain amount of resources allocated in each cell. Herders, represented as roles as: considerate or greedy, consume the resources in their neighborhood either considerately or greedily as shown in Figure~\ref{fig:payoffMatrices}c (payoffs they receive are the resources consumed in the environment). They can use available resource in their neighboring cells selected randomly.

The resources replenish with a constant natural growth rate of $c$. In addition, the growth rate can be increased by $w$ if a cell is occupied by a worker (otherwise $w=0$). Consequently, the growth of the resource in each cell is modeled by the logistic growth model as:
\begin{equation}
 \frac{\partial D^{(i,j)}}{\partial t} = (c + w)(1- \frac{D^{(i,j)}}{K})D^{(i,j)}
\end{equation}
where $K$ is the upper limit (carrying capacity) of the resources. Note that, when the resource in a cell is depleted $D^{(i,j)} = 0$, it cannot be recovered.

\section{Results}

\subsection{Social norms enable lifetime-learning of division of labor}

\begin{figure*}[!ht]
 \centering
 \includegraphics[width=0.99\textwidth]{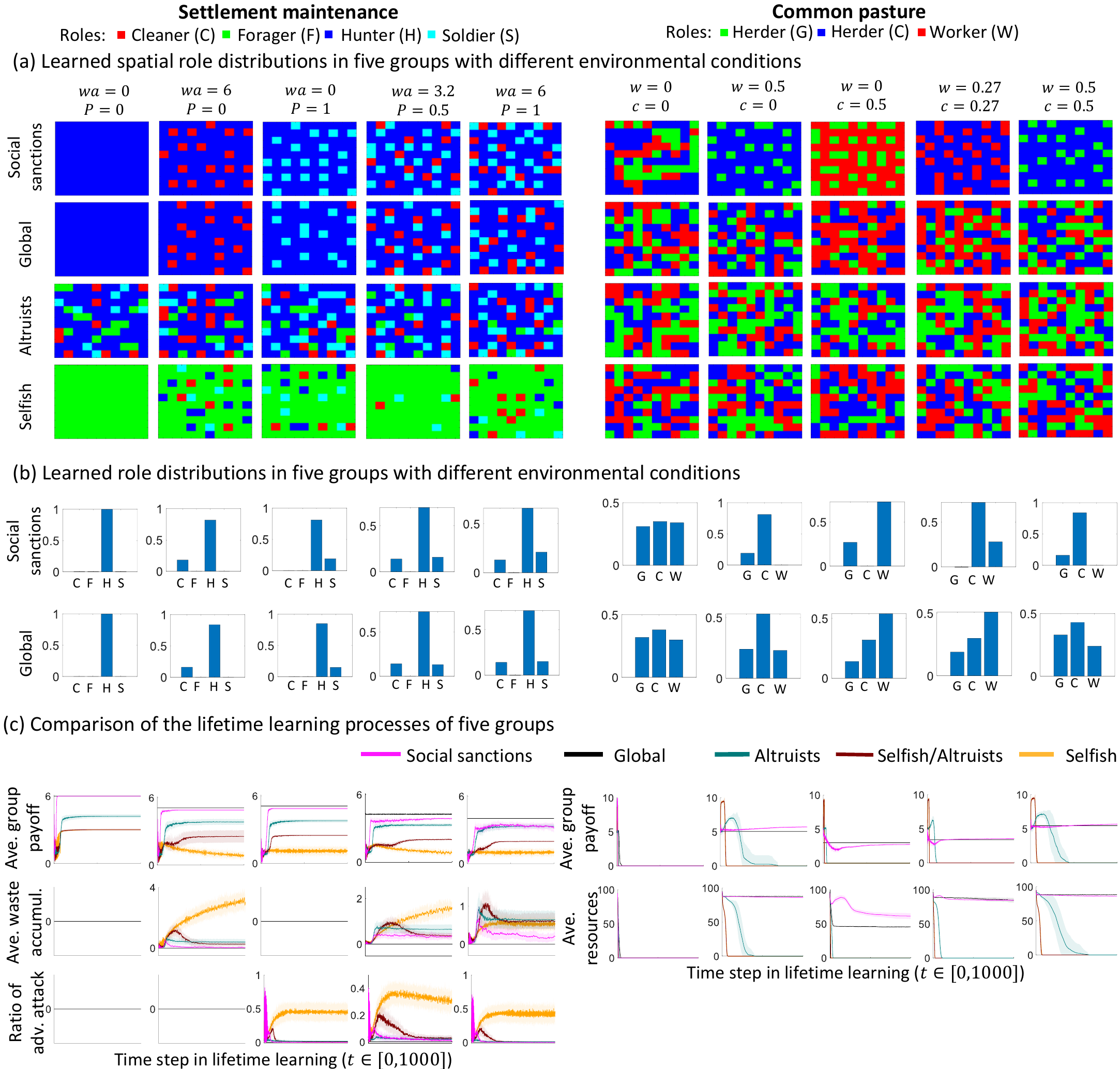}
 \caption{Social sanctions facilitate learning role distributions in self-interested lifetime-learning individuals various environment instances in settlement maintenance and common pasture games. The results shown in each row corresponds to a distinct environment parameter settings specified on top of each column, $wa$ (waste accumulation) and $P$ (adversarial attack probability) in settlement maintenance, and $w$ (worker growth rate) and $c$ (natural growth rate) in common pasture game. Shown in (a) and (b), different role distributions are learned locally (enforced by the decentralized social sanctions) to maximize group payoffs. (c) shows the change in the average group payoff and other environmental properties during the learning processes with social sanctions, and other three compared approaches. }
 \label{fig:results}
\end{figure*}

In both games, two environmental parameters are used for testing the emergence of learning proper role distributions in five different environment variations. In the settlement maintenance game (see Figure~\ref{fig:results}a), these parameters are based on waste accumulation and adversarial attack probability, and initialized as: $(wa, P) = \{(0,0), (6,0), (0,1), (3.2,0.5), (6,1)\}$. In the common pasture game, these parameters are natural growth rate and worker rate, and are initialized as $(w, c) = \{(0,0), (0.5,0), (0,0.5), (0.27,0.27), (0.5,0.5)\}$. For the two environmental parameters in both of games, we chose max value ranges and initialized five parameter assignment cases, four on the limits, and one in the middle of the parameter ranges. We ran the evolutionary processes independently and separately for each of these environment instances multiple times (i.e., 30 runs each) with complexity regularization $\lambda = 0.2$ to find social norms that could provide optimum leaning in each one. We observed that, due to the complexity regularization, multiple runs of the evolutionary processes converged to similar social norms that can provide (near-)optimum group welfare. Without the complexity regularization, we observed the emergence of various social norms. The analysis of the effect of complexity regularization is provided in Section~\ref{lab:complexityRegularization}, and examples of the social norms and the spatial role distributions they converged are provided in Supplementary Material.

In Figure~\ref{fig:results}a, examples of learned spatial role distributions in five environment variants in both games are shown. The individuals converged on different role distributions depending on the environmental conditions. Interestingly, the emerged role distributions show clear signs of neighborhood patterns where the roles that benefit their neighbors, distributed in various neighborhoods for minimizing their overlap.

In the settlement maintenance game, since there is no need for cleaner and soldier roles when there is no waste accumulation and adversarial attacks ($wa = 0, P = 0$), all individuals converged to hunter role to achieve the maximum payoff. When waste accumulation or adversarial attack probabilities is increased, some individuals converge to cleaner or soldier roles respectively to mitigate the adverse effects of these environments. When both waste accumulation and adversarial attack probability are increased, we see the emergence of both cleaner and soldier roles.

In the common pasture game, when the natural growth and worker rates are low, there is no possibility of sustaining the resources in the environment. Therefore, we see random role selection in this case (when $w=0$ and $c=0$). When one or both of these factors increase, various role distributions can be learned to collect as much resources as possible while sustaining the environment. For instance, when the natural growth rate $c$ is high, the number of (considerate) herders increases. Interestingly, an increase in the number of worker roles can also be observed even though they do not have any function when $w=0$. This is due to the fact that workers can function as empty placeholder cells that prevents additional farmers in the environment, and therefore avoid excess use of resources. When the natural growth rate is maximum, we can observe the appearance of some greedy herders.

In both games, self-interested lifetime-learning individuals without social sanctions, referred to as ``selfish'' from now on, do not obtain good performance. In the settlement game, all the individuals converge to select the forager role since it is easier to learn because it doesn't depend on coordination with others. The hunter role, which can provide higher rewards, depends on coordination within the neighborhood, so it is more difficult to discover by chance since at least five individuals would have to select it simultaneously. In the common pasture game, selfish individuals learn the greedy herder role because it has the highest immediate payoff. However, when they do, they deplete all the resources of the environment quite quickly. Both games have critical roles that do not provide any payoff on their own: cleaner and soldier in settlement maintenance and worker in common pasture. Selfish individuals will not select these roles since they do not provide any payoff. Thus, groups of selfish individuals cannot establish a socially advantageous division of labor.

The learning performance of self-interested lifetime-learning individuals with social sanctions, referred to as ``social sanctions'' from now on, and selfish individuals were compared with three additional approaches. These were: global, altruists and selfish/altruists. Briefly, in the global approach, the role distributions of the groups are directly optimized by evolutionary algorithms where the role of each individual in each location is defined when they are initialized, and remains fixed throughout their lifetime. In selfish/altruists, the individuals aim to maximize the average payoffs of self and neighboring individuals. In case of altruists, they aim to maximize the average payoffs of neighboring individuals, not including themselves. The details of these approaches are provided in Methods section.

Figure~\ref{fig:results}c compares all five approaches to one another. In both games, the global approach provides the upper bound. This is as expected since it can take into account global knowledge of the problem which allows making improvements on some roles independently of others in different locations by keeping them constant. In addition, this approach does not involve the costs of lifetime-learning that arise from trial and error. Results from the social sanctions approach were the closest to this upper bound. The selfish individuals learn to perform the role that maximizes their payoff. However, without the other regulatory roles in the groups, the environment degrades quickly leading to the worst performance. Altruists and selfish/altruists improve over selfish individuals but do not perform as well as social sanctions.

\subsection{Complexity regularization promotes convergent evolution of simpler social norms}\label{lab:complexityRegularization}

\begin{figure*}[!ht]
 \centering
 \includegraphics[width=0.99\textwidth]{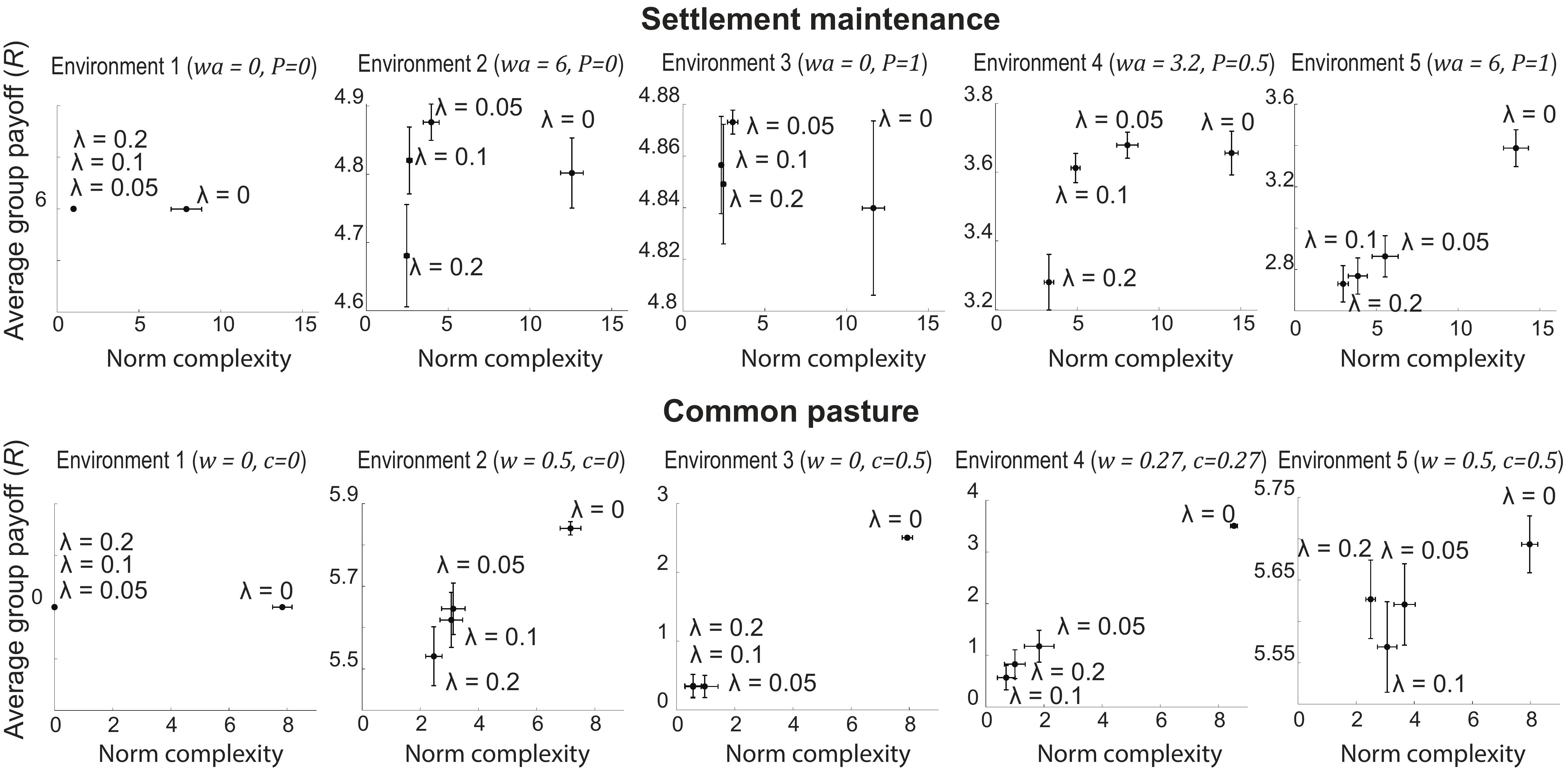}
 \caption{Complexity regularization promotes convergent evolution of simpler social norms. We observe that larger values of $\lambda$ lead to the emergence of social norms that are less complex. This is indicated by the results of the average group payoff versus their complexity in 5 different environments in settlement maintenance and common pasture problems depending on different $\lambda$ values. Each point shows the average of the result of 30 independent. Error bars show the standard error from the mean. }
 \label{fig:sensitivityAnalysisLambda}
\end{figure*}

Figure~\ref{fig:sensitivityAnalysisLambda} shows the relationship between the average group payoffs and the complexity of the norms for different complexity regularization ($\lambda$). To facilitate a better comparison, average group payoff shows only $R$ (see Equation~\eqref{eq:lifetimeReward}), therefore it does not include the complexity regularization ($-\lambda \lVert S \rVert_0$).
The complexity of the norms is assessed by the size of the social norms: the social norms that are composed of larger number of rules are more complex.

We observe that a larger $\lambda$ value leads to the emergence of simpler rules in all environments. Furthermore, there is no statistical difference (based on the Wilcoxon rank-sum test~\cite{wilcoxon1992individual} on $\alpha=0.05$ significance level) between the average payoff achieved with different $\lambda$ values in each environment except in environments 3 and 4 in common pasture. For the exception cases, cultural evolution processes with complexity regularization (when $\lambda>0$) was able to produce norms that can successfully promote the division of labor only in a limited number of independent processes (i.e. 4 and 7 out of 30 independent cultural evolutionary processes in Environments 3 and 4 respectively). Therefore, on average, the results are lower in those cases. If only cultural evolution processes which led to a successful division of labor are considered, then the results are statistically similar to the case without complexity regularization ($\lambda =0$).

Overall, complexity regularization is able to promote the emergence of norms that can achieve similar average group payoffs. Moreover, the complexity regularization promoted convergent evolution by means of producing similar norms at the end of independent cultural evolution processes (see Supplementary Material for examples of norms emerged as a result of the independent cultural evolution processes for $\lambda$ values 0 and 0.2). These results are further elaborated in the next two sub-sections.

\noindent{}\textbf{Settlement maintenance}

In \textbf{Environment 1}, independent of $\lambda$ value, social norms allow groups to achieve maximum average payoff of 6 (achieved when all individuals learn to choose Hunter role). When there is no importance given to the complexity regularization ($\lambda=0$), the average number of rules of evolved social norms is more than 8. Only a few of these evolved social norms converged to a similar rule set. The rest is converged to a unique and complex set of rules. On the other hand, when the importance to complexity introduced, all the norms converge to two possible types that consist of a single rule: Either, ``\textit{Hunter ought to discourage Forager by $X$ amount}'' or ``\textit{Forager ought to encourage Hunter by $Y$ amount}''. And here, $X$ and $Y$ can take values from 0.03 to 2.8; though discouragement is associated with a negative sign. This is quite intuitive, and in this case, encouragement/discouragement between these two roles allow them to learn to select Hunter to receive a higher payoff by cooperating with others.

Environments from 2 to 5 are more complex in that they require additional roles in the groups to achieve a higher average payoff. In case of Environment 2, Cleaner, in Environment 3, Soldier, and in 4 and 5, both Cleaner and Soldier roles are needed. In these environments, increasing $\lambda$ leads to the emergence of similar rules in multiple independent cultural evolutionary processes. In these environments, we notice the emergence of norms that are concerned with these roles in addition to the baseline norm that emerged in Environment 1. An example of this is a norm emerged for Environment 2, when $\lambda = 0.2$: ``\textit{(1) Hunter ought to discourage Forager by -1.45, and (2) Hunter ought to encourage Cleaner by 0.60}''. The first part of this norm is the same as the norms that emerged for Environment 1, in addition, second part emerged to encourage Cleaner role in the group. Another example is for Environment 4: ``\textit{(1) Hunter ought to discourage Forager by -2.28, and (2) Hunter ought to encourage Cleaner by 0.57, and (3) Hunter ought to encourage Soldier by 0.42}''. Here, we observe appearance of two rules regarding Cleaner and Soldier role in addition to the norm emerged in Environment 1. Similar patterns can be observed in the results for other environments provided in Supplementary Material in detail.

Without complexity regularization, while the difficulty of the environments increases, the complexity of the emerged norms also increases. For example, complexity remains around 8 in Environment 1. This shows that through the cultural evolution processes the size of the norms remains stable (considering expected norm size during the initialization is 8). However, in other environments, the complexity of the norms increases throughout the processes. This indicates that larger number of rules may be required to perform better in those environments. Nevertheless, when the complexity regularization is introduced, especially when $\lambda=0.2$, the average sizes of norms are reduced around 2.5 in environments 2 and 3, and around 3 and 3.5 in environments 4 and 5 respectively.

In \textbf{Environment 2}, the simplest rules have the length of 2. Interestingly, these rules are composed of the same rule that is emerged for Environment 1 to promote the cooperation between Hunter roles, and an additional rule that promotes learning Cleaner roles. Similar outcome is observed for \textbf{Environment 3} where Soldier roles are needed on top of the cooperation between Hunter roles to achieve the optimum average payoff. Finally, in \textbf{Environments 4 and 5}, the lengths of the emerged rules range from 3-5. Similarly, here, we observe appearance of the conjunction of the rules emerged in Environments 1, 2 and 3 since both Cleaner and Soldier roles are needed in this environment.

\noindent{}\textbf{Common pasture}

It is not possible to achieve any payoff in \textbf{Environment 1} because of the parameters of the environment since they do not allow natural and worker growth rate for recovering the environmental resources. Therefore the norms that are found at the end of the cultural evolutionary processes are random. However, when the complexity regularization is introduced, the length of the rules converged to 0. That is because, if no norm is helpful, having no norm is the best.

In \textbf{Environment 2}, we observe convergent evolution of the same type of social norm that is composed of two rules where Greedy (Herder) encourages Considerate (Herder), and Considerate (Herder) encourages Considerate (Herder). This norm facilitates the emergence of role distributions that consist mainly of Considerate (Herders) and some Greedy (Herders) that are spread around. This pattern leads to the exploitation of the environmental resources with Greedy (Herder) without disrupting the sustainability.

In \textbf{Environment 3}, only four independent cultural evolution processes produced social norms that led to sustainable use of environmental resources. These social norms are composed of partially different rules although they have similar length and produce similar division of labor patterns exception in one case (provided in Supplementary Material). Similarly, in \textbf{Environment 4}, only seven of the processes produced social norms that promote sustainable use of resources. These rules are composed of partially the same rules. In \textbf{Environment 5}, in many cases, the norms are composed of two rules and converged to the same rules.

\section{Discussion}

Human ecological success has been underpinned by our species' ability to overcome the challenges of group collaboration. In these situations, the individuals who compose a group often need to divide labor and perform various specialized roles collaboratively. However, in groups consisting of self-interested individuals that aim to maximize their own payoff, establishing such collaboration may not be straightforward. In this work, we focused on the problem of learning such a division of labor in groups of self-interested lifetime-learning individuals. We modeled social norms as decentralized social sanctioning patterns and studied how they can encourage individuals to cooperate. We also proposed a regularized cultural group evolution model to study how such norms which enable the learning of prosocial roles can be established. We demonstrated this learning problem, and how our model of social norm evolution resolves it, for two different spatial games. In both cases the emergent social norms discovered by our cultural group evolution model were successful in incentivizing individuals to learn to collaborate with one another.

There is an extensive literature on the evolution of cooperative strategies in populations of self-interested individuals~\cite{axelrod1981evolution,axelrod1986evolutionary,nowak2006five,riolo2001evolution}. Some, like us, studied the emergence of cooperation in spatial game settings~\cite{nowak1992evolutionary,nowak1994spatial,szolnoki2012conditional}. However, this line of work usually does not model lifetime-learning. We modeled individual-level lifetime-learning in conjunction with group-level norm evolution. In our model, the two levels were linked because group-level norms guide the lifetime-learning of individuals. At the same time, whether a newly innovated norm undergoing testing makes the cut and becomes established depends on its guiding the learning of all individuals to a more socially advantageous outcome. Wang et al. (2019) took an analogous two-level strategy, integrating individual reinforcement learning with an evolutionary process, which in their case, determined the individuals' reward functions~\cite{wang2019evolving}. They showed that group selection was needed to get reward functions that incentivize altruistic behavior to evolve, a result that supports a key assumption of our model: that social norms evolve on the level of groups.

Most evolutionary game theory-based models share a common assumption that individuals make one choice and stick with it throughout their lives~\cite{weibull1997evolutionary}. While others model the evolution and spread of successful strategies~\cite{traulsen2009exploration,yaman2022meta} through social learning over a population of individuals. In contrast, the cultural evolutionary process in our model operates on the social norm level. Additionally, individuals are endowed with the capability of changing their behavior based on their own experience. Therefore, we can assume that individuals can re-select what role to play over and over again throughout their lives. As a result, our model is more about deciding what you want to do today than it is about deciding what to do with your life. It should correspond best to situations where roles may be revisited each day. There are likely many such roles in a foraging society. For instance, you could hunt one day and gather the next day. In contrast, the evolutionary game theory approaches where role definition is specified genetically (since this can make it irrevocable) may be better suited to modeling situations faced by modern people who must decide which of several highly specialized careers to pursue e.g. doctor or lawyer. In these situations individuals find it difficult to switch roles after investing considerable time into their training. This is why results from studies of ostensibly similar games to those we study, where a surplus may be obtained if all roles are filled but some roles pay better than others, have been interpreted in the evolutionary game theory context as speaking to the emergence of social stratification and inequality between social classes or castes~\cite{henrich2008division}. However, even though we allow individuals in our model to change roles on every time step, we still find that reinforcement learning converges to a stable choice of roles. So our model could still be used to study social stratification and inequality. From this lens, it is worth pointing out that, unlike models such as \cite{henrich2008division}, we do not require individuals to engage in social learning, nor do we need to posit accumulating differences in social capital between groups as assumed by other models (e.g.~\cite{lundberg1998persistence}). In our model, agents only learn from their own direct experience in different roles and from normative sanctioning, but this is still sufficient for stratification to emerge at convergence.

The game-theoretic literature on social norms has pursued two broad categories of mechanisms by which norms may affect cooperation. They are (A) transforming the payoffs (via sanctioning or equivalent) into a new effective game with different Nash equilibria, e.g. making mutual cooperation an equilibrium in a transformed game derived from the Prisoners' Dilemma~\cite{ullmann1977emergence, kelley2003atlas}, and (B) equilibrium selection. When social norms are regarded as equilibrium selection devices, the assumption is that many Nash equilibria are possible and would be stable if achieved, and that some of them are quite bad (e.g. tragedy of the commons situations). Society must navigate its way to the good equilibria while avoiding the bad ones. In this line of work, the norm is regarded as a piece of public knowledge that individuals may condition their behavior on in order to rationally coordinate with one another~\cite{vanderschraaf1995convention, gintis2010social, hadfield2012law, axtell2000emergence}. For instance, a focal point is a ``default choice'' from which it would be irrational to deviate when you know others are trying to coordinate~\cite{schelling1960strategy}. Many models incorporate aspects of both mechanisms (e.g.~\cite{bicchieri2006grammar, boyd2021arbitration}). Our approach resonates more strongly with approach (A). The two games we studied, settlement maintenance and commons pasture, are social dilemmas. That is, they contain socially deficient Nash equilibria, and purely egoistic, utility-maximizing agents do not learn to cooperate; they never take on the unpaid-but-critical roles. In our model, social norms are patterns of social sanctioning that affect the rewards individuals experience. We can see them as transforming the game into a different one where it is profitable to select the unpaid-but-critical roles because other individuals positively sanction that choice, effectively paying the agents who choose them for the service they provide to the group. It is thus critical in our model that the norms themselves do not evolve within individuals, and individuals cannot unilaterally decide whether to sanction or not. Norms must evolve on the level of the cultural group, using a ``fitness'' that incorporates the well-being of all the group's individuals, otherwise they could not incentivize individuals to choose dominated strategies, as they must to achieve a stable and socially advantageous division of labor~\cite{henrich2004cultural}.

\section{Methods}\label{sec:methods}

\subsection{Social norm representation and evolution}

Social sanctioning rules are set of statements that can be written as follows:

\begin{equation}
  \text{Social sanctioning rules} :
    \begin{cases}
      (1) &\text{``$\rho_1$ ought to encourage $\rho_1$ by $s_{11}$''}\\
      (2) &\text{``$\rho_1$ ought to discourage $\rho_2$ by $s_{12}$''}\\
      (..) & \text{......}\\
      (k\times k) &\text{``$\rho_k$ ought to encourage $\rho_k$ by $s_{kk}$''}\\
    \end{cases}
\end{equation}
where $\{\rho_1, \rho_2, \hdots, \rho_k\}$ are $k$ number of roles each individual can choose and $s_{ij}$ are the amount of sanctioning applied between the individuals that perform these roles. The rules are implemented from the \textit{agent-centric} point of view, where each individual implements the sanctioning value to the other individuals within their social networks (neighboring individuals). Rewards and punishments that used for encouragements and discouragements.

Since there are $k$ number of roles, it is possible to define $k\times k$ number of rules. For computational reasons, we represent these rules in form of a matrix where the cells indicate the amount of sanction, and rows and columns represent the roles that the social sanction is imposed (see Figure~\ref{fig:socialSanctions}).

For the cultural evolutionary process of the rules, we use three operators for perturbing existing social sanctioning rules to generate new ones. As discuss in Section~\ref{sec:culturalEvoSocialNorms}, these operators are: Gaussian perturbation, rule addition and rule deletion. These operators perform changes on the matrix representation of the rules where Gaussian perturbation is applied to the non-zero cells with a probability we refer to as mutation probability ($mp$). Furthermore, Gaussian perturbation has another parameter called mutation rate ($mr$) to adjust the standard deviation of the perturbation. After this change, we make sure that the values stays within the boundary range. Rule addition and deletion operators add or delete a rule (by adding a sanctioning value to a cell with zero value or assign zero value to a cell with non-zero value) with rule addition probability ($rap$) and rule deletion probability ($rdp$) respectively. Algorithm~\ref{alg:rulePerturbation} shows the perturbation process of the social sanctioning rules. Due to the stochasticity of the perturbation process, we make sure that at least one perturbation operator is triggered so that there could at least be one form of change in the social sanctioning rules.

\begin{algorithm}[!ht]
    \begin{algorithmic}[1]
	    \Procedure{PerturbNorm}{$\boldsymbol{S},mp$,$mr$,$rap$, $rdp$}
                \If{$rand < mp$}  \Comment{Gaussian perturbation for the amount of sanctioning}
                    \State $s_{ij}\gets s_{ij} + \mathcal{N}(0,mr), \forall s_{ij} \in \boldsymbol{S} \text{ such that } s_{ij} \neq 0$	           
                \EndIf
	        \If{$rand < rad$} \Comment{Rule addition}
                    \State $(i,j)\gets rand \text{ such that } s_{ij} = 0$ 
                    \State $s_{ij}\gets \mathcal{N}(0,1)$           
                \EndIf
                \If{$rand < rdp$} \Comment{Rule deletion}
                    \State $(i,j)\gets rand \text{ such that } s_{ij} \neq 0$ 
                    \State $s_{ij}\gets 0$           
                \EndIf
                \Return $\boldsymbol{S}$
		\EndProcedure
	\end{algorithmic}
\caption{Perturbation process a social norms to generate a new one. }
\label{alg:rulePerturbation}
\end{algorithm}

\subsection{Social sanctioning procedure}\label{sec:socialSanctioning}

The update procedure of the rewards based on decentralized social sanctioning mechanism is shown in Algorithm~\ref{alg:rewardUpdateSocialSanctions}. Here, for each neighbor of each individual (in random order), the social sanction (defined in the social sanctioning matrix) is imposed based on the role selection of the focal individual and its neighbor. We used also a version where the social sanctions are imposed based on a specified order of the neighbors (such as starting always from the neighbors on the upper right to the neighbor on the lower left). However, the random order is preferred to avoid the dependency to this ordering during the sanctioning process.

\begin{algorithm}[!ht]
    \begin{algorithmic}[1]
	    \Procedure{UpdateRewards}{}
	    \ForEach{$i\in I$} \Comment{In random order}
	        \State $neighbors = $getNeighbors$(i)$
	        \ForEach{$j\in neighbors$} \Comment{In random order}
	            \State $s_{ij} = $sanction$(role(i,t), role(j,t))$
	            \If{$s_{ij} > 0$ and $s_{ij} \geq r_t^{(i)}$} \Comment{Individual i is rewarding j}
	                \State $r_t^{(i)} = r^{(i)} - s_{ij}$
	                \State $r_t^{(j)} = r^{(j)} + s_{ij}$
	            \ElsIf {$s_{ij} < 0$ and $|s_{ij}| \geq r_t^{(j)}$} \Comment{Individual i is punishing j}
	                \State $r_t^{(i)} = r_t^{(i)} + |s_{ij}|$
	                \State $r_t^{(j)} = r_t^{(j)} - |s_{ij}|$
	            \EndIf
	        \EndFor
	    \EndFor
	    
		\EndProcedure
	\end{algorithmic}
\caption{The update procedure of the individual rewards based on the decentralized social sanctioning.}
\label{alg:rewardUpdateSocialSanctions}
\end{algorithm} 

\subsection{Social norm evolution}
Algorithm~\ref{alg:normEvolution} provides the evolutionary process of the social norms. The algorithm starts with a randomly initialized norm (by randomly initializing the cells of the social sanctioning matrix) and aims to find the norm that provides the highest average group payoff. This is achieved by making a small change in the norm (via Gaussian perturbations, adding or removing rules) and testing whether the new norm provides better performance (learning of better role distributions that can maximize the group payoff). If so, the previous norm is replaced by the new, better performing norm. This process is repeated until a maximum number of iteration is reached. The social norm at the final iteration is then provided as the result of the algorithm.

\begin{algorithm}[ht!]
    \begin{algorithmic}[1]
	    \Procedure{NormEvolution}{}
     \\ \textit{$\lambda$: weight for complexity.}
     \\ \textit{$mp,mr,rap,rdp$: mutation probability, mutation rate, rule addition probability, rule deletion probability.}
	        \State $\boldsymbol{S} \gets \text{randomSocialNorm}()$
	        \State $Ft \gets \text{EVAL}(\boldsymbol{S}) - \lambda || \boldsymbol{S} ||_0$
	        \State $i =1$ \Comment{counter for the number of iterations}
            \While {$i < maxIter$} \Comment{until the max iterations}
                    \State $\boldsymbol{S}^{\prime} \gets \text{PerturbNorm($\boldsymbol{S},mp,mr,rap,rdp$)}$
                    \State $Ft^{\prime} \gets \text{EVAL}(\boldsymbol{S}^{\prime}) - \lambda || \boldsymbol{S} ||_0$
                    \If {$Ft^{\prime} > Ft$}
                        \State $\boldsymbol{S} \gets \boldsymbol{S}^{\prime}$
                        \State $Ft \gets Ft^{\prime}$
                    \EndIf
                    \State $i\gets i + 1$
            \EndWhile 
		\EndProcedure
	\end{algorithmic}
\caption{Pseudocode for the evolution of social norms.}
\label{alg:normEvolution}
\end{algorithm}

\subsection{Other compared learning approaches}

\subsubsection{Global}

For the global approach, instead of optimizing the social norms, we optimize the spatial role distributions directly using standard genetic algorithms~\cite{eiben2003introduction}. The pseudocode for the global approach is provided in Algorithm~\ref{alg:centralized}. 

For both games, we aim to find a 10 by 10 matrix where each cell can take one of the roles defined for each game. In this case, since there are 4 and 3 possible role selections in each cell, there are $4^{100}$ and $3^{100}$ possible spatial role distributions in the settlement maintenance and common pasture games, respectively. 

\begin{algorithm}[ht!]
    \begin{algorithmic}[1]
	    \Procedure{Global}{}
	        \State $\boldsymbol{X} \gets \text{randomSetRoleDistributions}()$
	        \State $F \gets \text{Evaluate}(\boldsymbol{X})$
	        \State $g=1$ \Comment{counter for the number of iterations}
            \While {$g < maxGen$} \Comment{until the max number of generations}
                    \State $\boldsymbol{X}^{\prime} \gets \text{Crossover}(\boldsymbol{X})$
                    \State $\boldsymbol{X}^{\prime\prime} \gets \text{Mutate}(\boldsymbol{X\prime})$
                    \State $F^{\prime} \gets \text{Evaluate}(\boldsymbol{X}^{\prime\prime})$
                    \State $[\boldsymbol{X}, \boldsymbol{F}] \gets \text{Select}(\boldsymbol{X}\cup \boldsymbol{X}^{\prime\prime}, F\cup F^{\prime}) $
                    \State $g\gets g + 1$
            \EndWhile 
		\EndProcedure
	\end{algorithmic}
\caption{Pseudocode for optimizing the spatial role distributions directly using the global approach.}
\label{alg:centralized}
\end{algorithm}

Using genetic algorithms, we represent a set of randomly initialized role distributions $\boldsymbol{X}$ where each role distribution in the set represents a candidate solution. We encode each candidate solution as a vector form (1 by 100), flattening the corresponding 10 by 10 spatial role distributions for computation convenience. The goal is to search this solution space to find the solution (i.e., the role distribution) that maximizes the average group payoff. We do that by applying evolutionary inspired operators (\textit{crossover} and \textit{mutation}) to produce an offspring candidate solution set $\boldsymbol{X}^{\prime\prime}$ using the current set of candidate solutions. Then, we create a new set of solutions by selecting the best solutions from the union of the current $\boldsymbol{X}$ and offspring sets $\boldsymbol{X}^{\prime\prime}$. This process is performed iteratively until a certain number of generations. At the end, the algorithm provides the best solution that is the spatial role distribution that provides the maximum average group payoff encountered during the search process. The evaluation process of each candidate solution does not involve lifetime-learning since the roles of the individuals in each cell are directly specified. To evaluate the solution, we run the game process for a certain number of action steps, record the rewards achieved by each individual, and then find the average reward of the group. 

\subsubsection{Selfish}

Selfish individuals are modelled based on a reinforcement learning approach, known as the $\epsilon$-greedy algorithm~\cite{sutton2018reinforcement}, that aims to maximize their own payoffs based on the Equations~\eqref{eq:roleSelection} and \eqref{eq:ILrewardUpdate}~\cite{sutton2018reinforcement}. This is the basic form depicted as the \textbf{Stage I} in Figure~\ref{fig:socialSanctions} without applying the social sanction part shown as the \textbf{Stage II}.

\subsubsection{Selfish/altruists and altruists}

Selfish/altruists individuals aim to maximize the average of the payoffs of their own and the average payoffs of the other individuals in the neighborhood, and altruist individuals aim to maximize the average payoffs of the individuals in their neighborhood. To model these behaviors, we modified the equation used  by selfish individuals (given in Equation~\eqref{eq:ILrewardUpdate}) as follows:
\begin{equation}
\label{eq:altruists}
Q^{(i)}_{t+1}(A^{(i)}_{t}) = Q^{(i)}_{t}(A^{(i)}_{t}) + \alpha \left[r - Q^{(i)}_{t}(A^{(i)}_{t}) \right]
\end{equation}
\begin{equation}
\label{eq:altruistsRew}
r = \Omega\frac{1}{N}\sum r^{(N)}_t + (1-\Omega)r^{(i)}_t
\end{equation}
where $\Omega\in \{0.5, 1\}$ is a parameter that specifies the weight given to maximizing the average payoffs of the individuals that are in the neighborhood (denoted as $\frac{1}{N}\sum r^{(N)}_t$, where $N$ is the number of individuals within the neighborhood), rather than an individual's own payoff. For selfish/altruist behavior, we set $\Omega = 0.5$, while for altruist behavior $\Omega = 1$. Note that, when $\Omega=0$, we achieve selfish behavior.


\appendix

\newpage

\section{Sensitivity analysis on the learning rate ($\alpha$)}\label{lab:sensitivityAnalysisAlpha}

\begin{figure*}[!ht]
\centering
\includegraphics[width=0.95\textwidth]{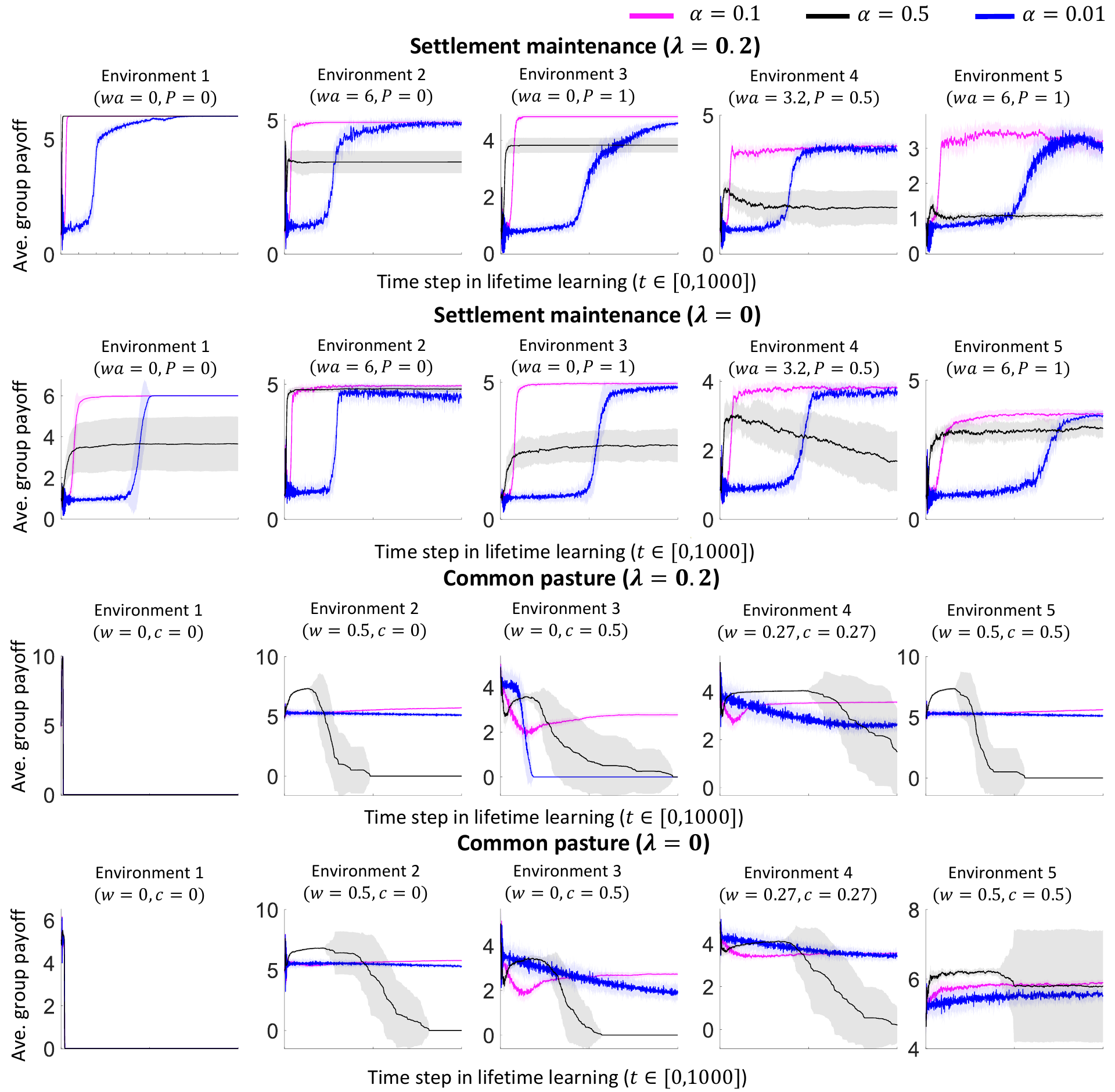}
\caption{How sensitive is the performance of the emerged social norms when ($\alpha = 0.1$) to low or high learning rates ($\alpha = 0.01$ and $\alpha = 0.5$)? In most cases, a lower learning rate ($\alpha = 0.01$) leads to similar results with a slower learning speed. On the other hand, in most cases, a higher learning rate ($\alpha = 0.5$) does not lead to a better/similar performance. This may be due to a greedy learning process where the individuals get stuck on a suboptimal role distribution without much exploration. }
\label{fig:sensitivityAlpha}
\end{figure*}

\begin{figure*}[!ht]
\centering
\includegraphics[width=0.95\textwidth]{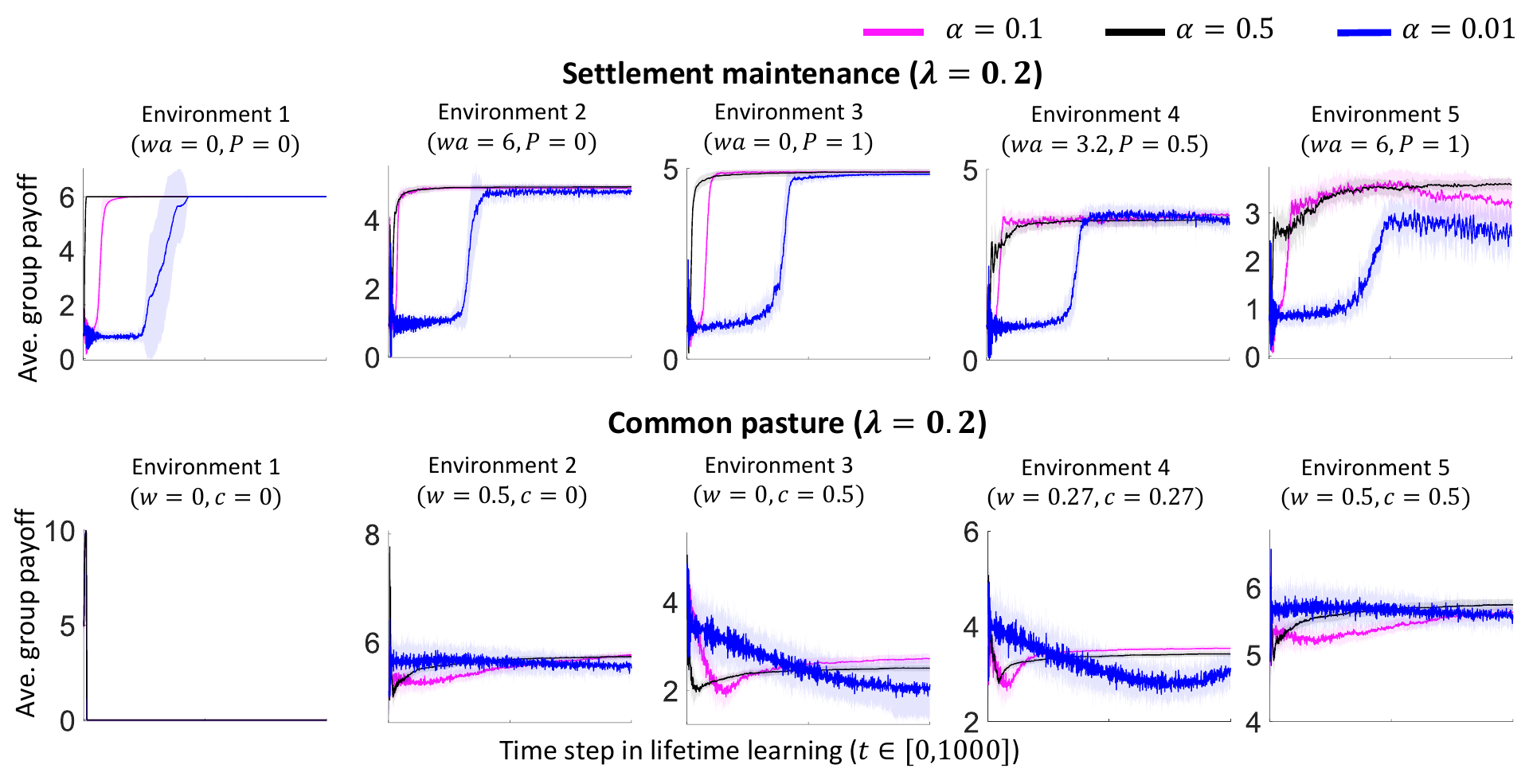}
\caption{when social norms are evolved particularly for low and high learning rates, they can produce similar results.}
\label{fig:sensitivityAlphaOptimized}
\end{figure*}

We tested the performance of the social norms evolved for learning rate ($\alpha=0.1$) on the learning processes with high ($\alpha=0.5$) and low ($\alpha=0.01$) settings. The results are shown in Figure~\ref{fig:sensitivityAlpha}. Based on the results, social norms that are evolved for $\alpha=0.1$ lead to similar but slow learning processes in case of low learning rate, however, they do not lead to similar or better performance when the learning rate is high. On the other hand, when social norms are evolved particularly for low and high learning rates, shown in Figure~\ref{fig:sensitivityAlphaOptimized}, we observe the emergence of social norms that can produce similar results.

\newpage

\section{Sensitivity analysis on the parameters of the cultural evolution}\label{lab:sensitivityAnalysis}

\begin{figure*}[!ht]
\centering
\includegraphics[width=0.95\textwidth]{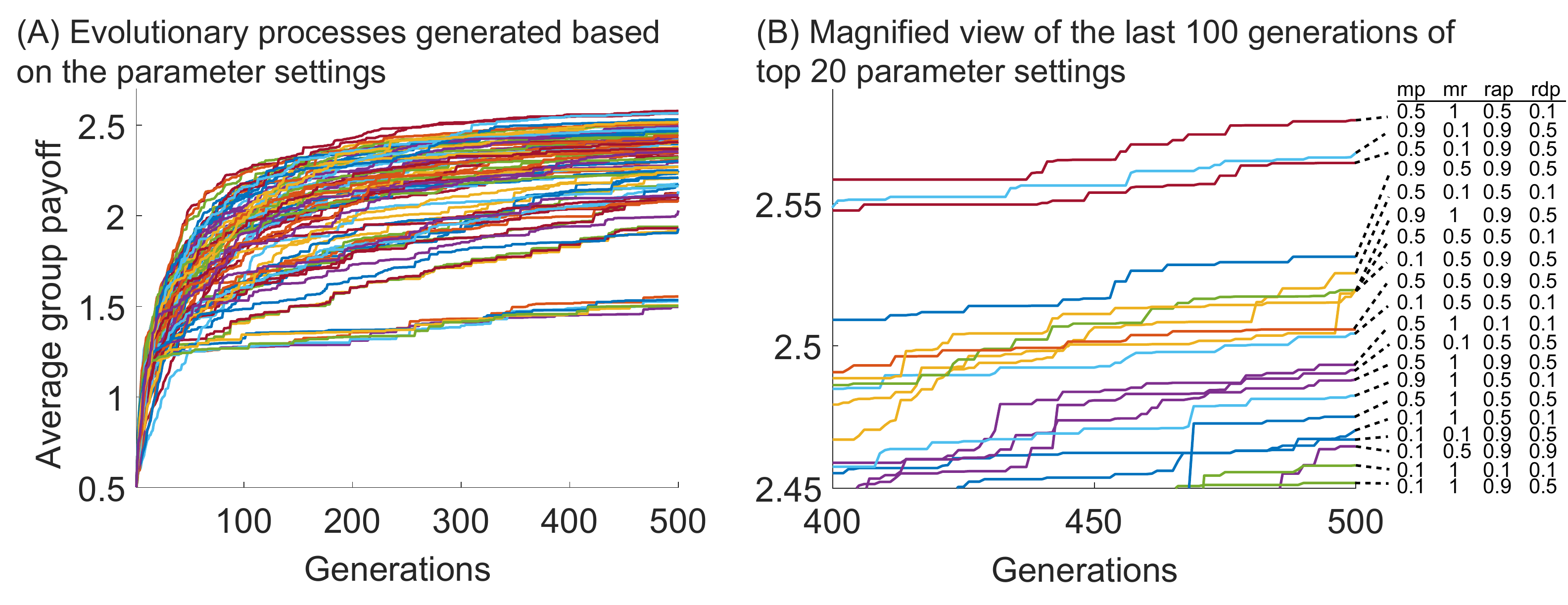}
\caption{Sensitivity analysis of the evolutionary process in respect to the parameters used to produce social sanctioning rule variations. Parameters used for generating variation in social sanctioning rules: mutation probability ($mp$), mutation rate ($mr$), rule addition probability ($rap$), rule deletion probability ($rdp$). Overall, parameters that modify sanctioning values with a lower probability and higher variance (e.g. $mp \le 0.5$ and $mr \ge 0.5$) or higher probability and lower variance (e.g. $mp \ge 0.5$ and $mr \le 0.5$), and high probability of adding new rules (e.g. $rap \ge 0.5$), and low probability of deleting rules (e.g. $rdp \le 0.5$) achieved higher average group payoffs at the end of the evolutionary processes.}
\label{fig:sensitivityEvolutionPars}
\end{figure*}

In Figure \ref{fig:sensitivityEvolutionPars}, we show the results of evolutionary processes that use different sets of parameters to adjust the intensity of producing variation in social sanctioning rules. We define three parameters (to correspond to low, medium and high values) to test the effect of four variables: mutation probability ($mp = \{0.1, 0.5, 0.9\}$), mutation rate ($mr=\{0.1, 0.5, 1\}$), rule addition probability ($rap = \{0.1, 0.5, 0.9\}$) and rule deletion probability ($rdp = \{0.1, 0.5, 0.9\}$). All the combinations of these parameters yields to 81 independent social sanctioning rule evolution processes. We ran 30 independent evolutionary processes for all 81 parameter combinations on the settlement maintenance game with $wa=6$ and $P=1$ settings.

Firstly, although, the parameter settings are ranked based on the average group payoff they received over these 30 runs (see Figure \ref{fig:sensitivityEvolutionPars} (B)), top performing parameter setting does not show significantly different results relative to the following top 20 best ranked parameter settings.

\newpage

\section{Evolved social norms and the effect of the complexity regularization}\label{lab:evolvedSocailNorms}

\begin{figure*}[!hb]
\centering
\includegraphics[width=0.95\textwidth]{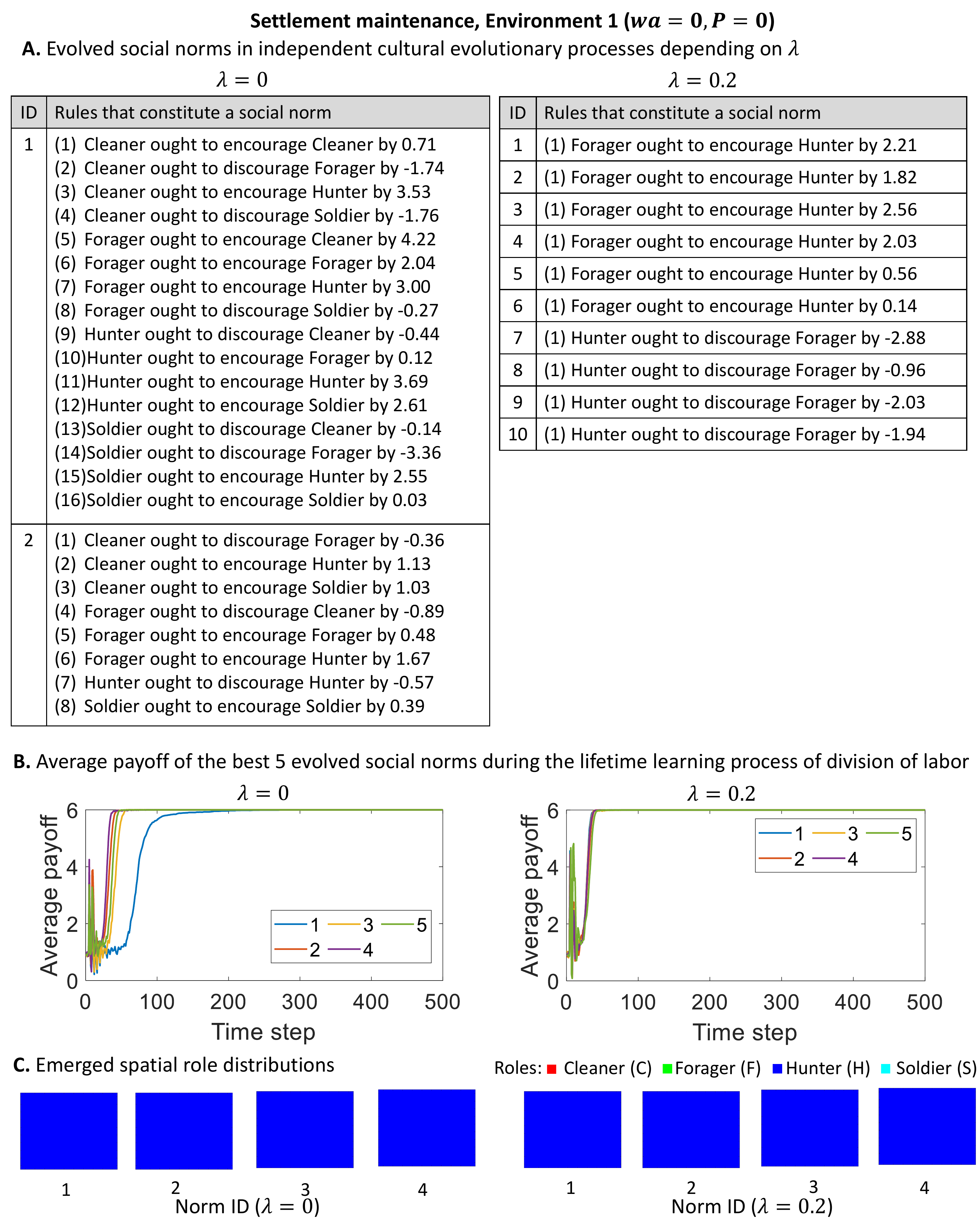}
\label{fig:Env1Settlement}
\end{figure*}

\begin{figure*}[!hb]
\centering
\includegraphics[width=0.95\textwidth]{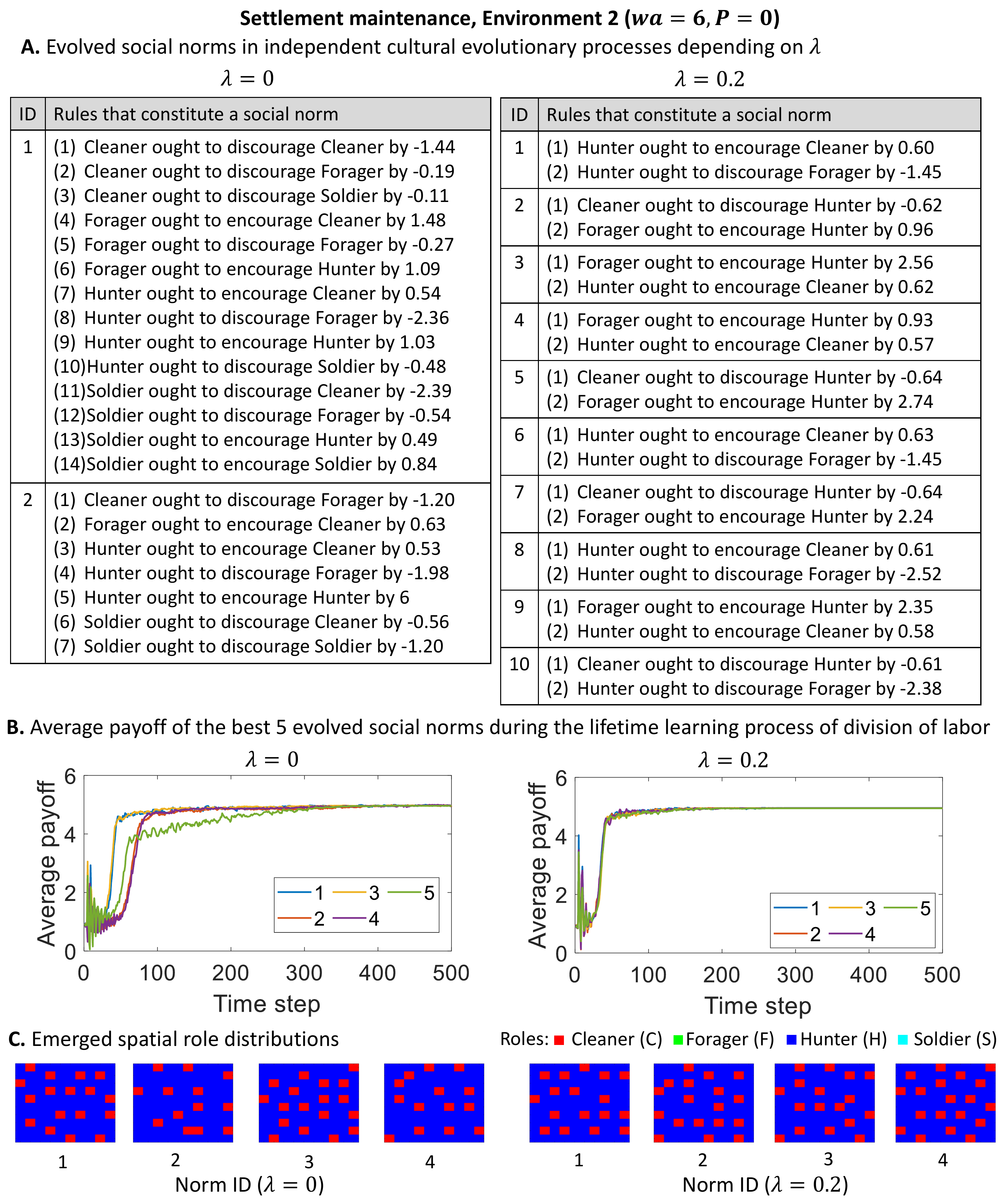}
\label{fig:Env2Settlement}
\end{figure*}

\begin{figure*}[!hb]
\centering
\includegraphics[width=0.95\textwidth]{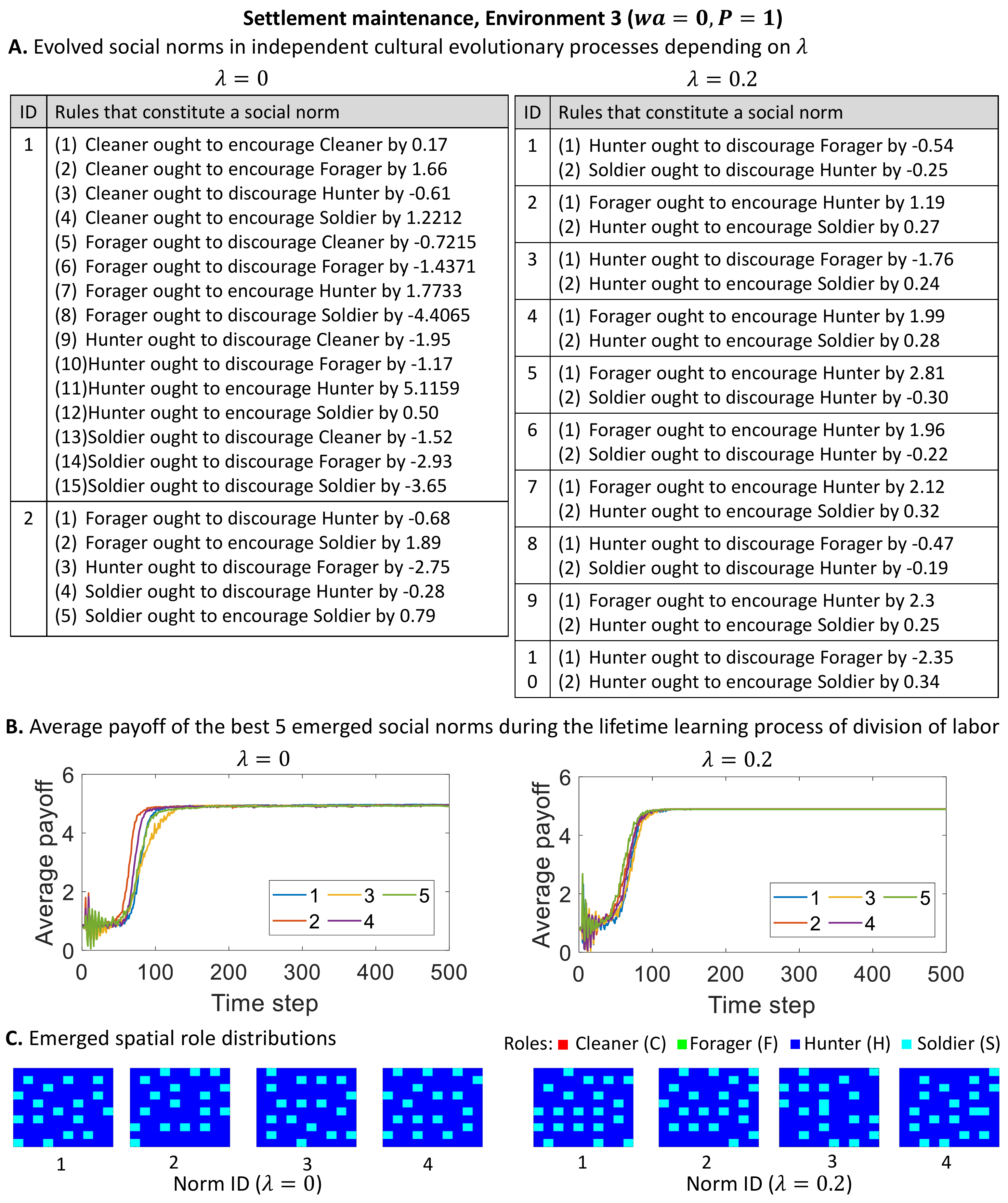}
\label{fig:Env3Settlement}
\end{figure*}

\begin{figure*}[!hb]
\centering
\includegraphics[width=0.95\textwidth]{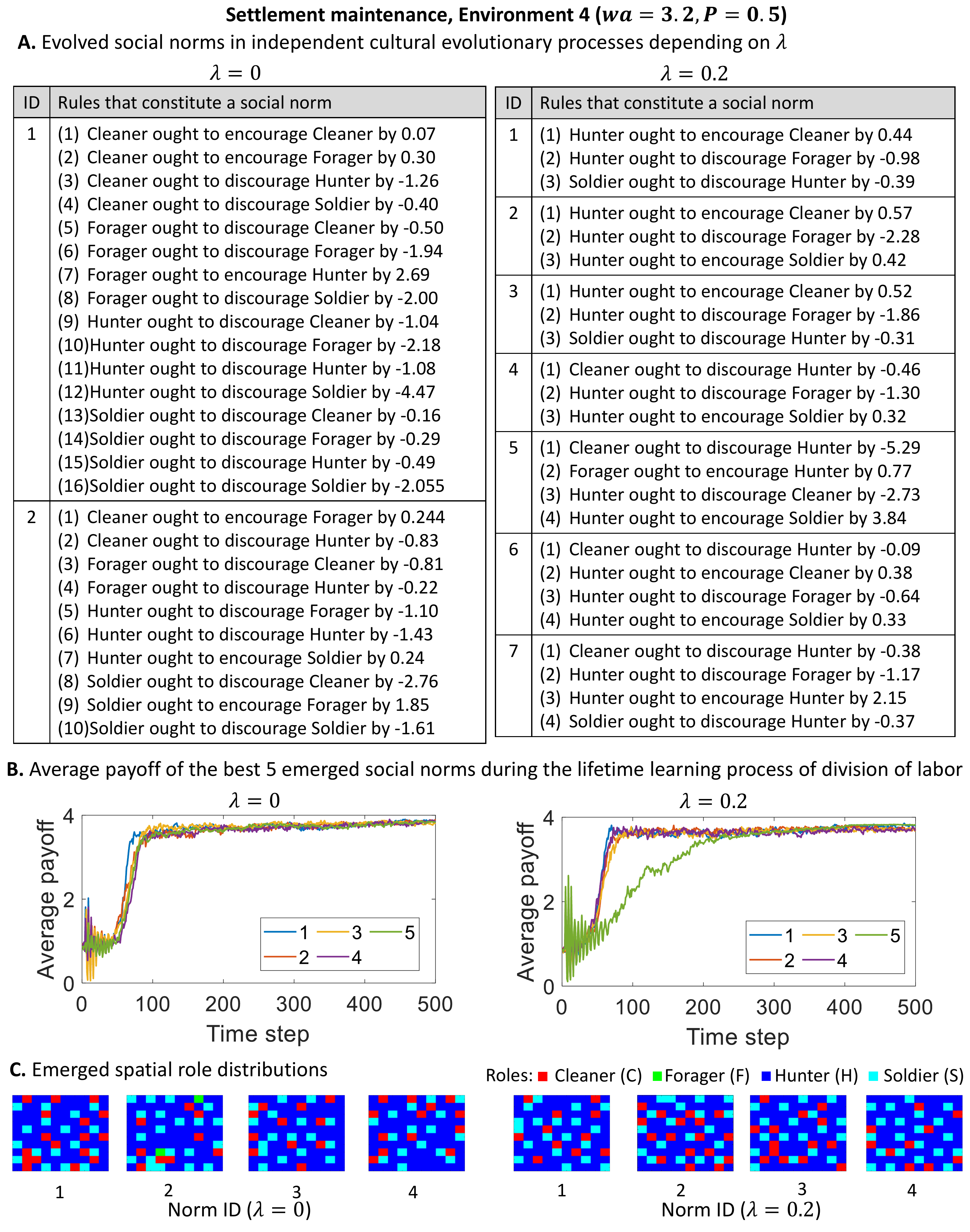}
\label{fig:Env4Settlement}
\end{figure*}

\begin{figure*}[!hb]
\centering
\includegraphics[width=0.95\textwidth]{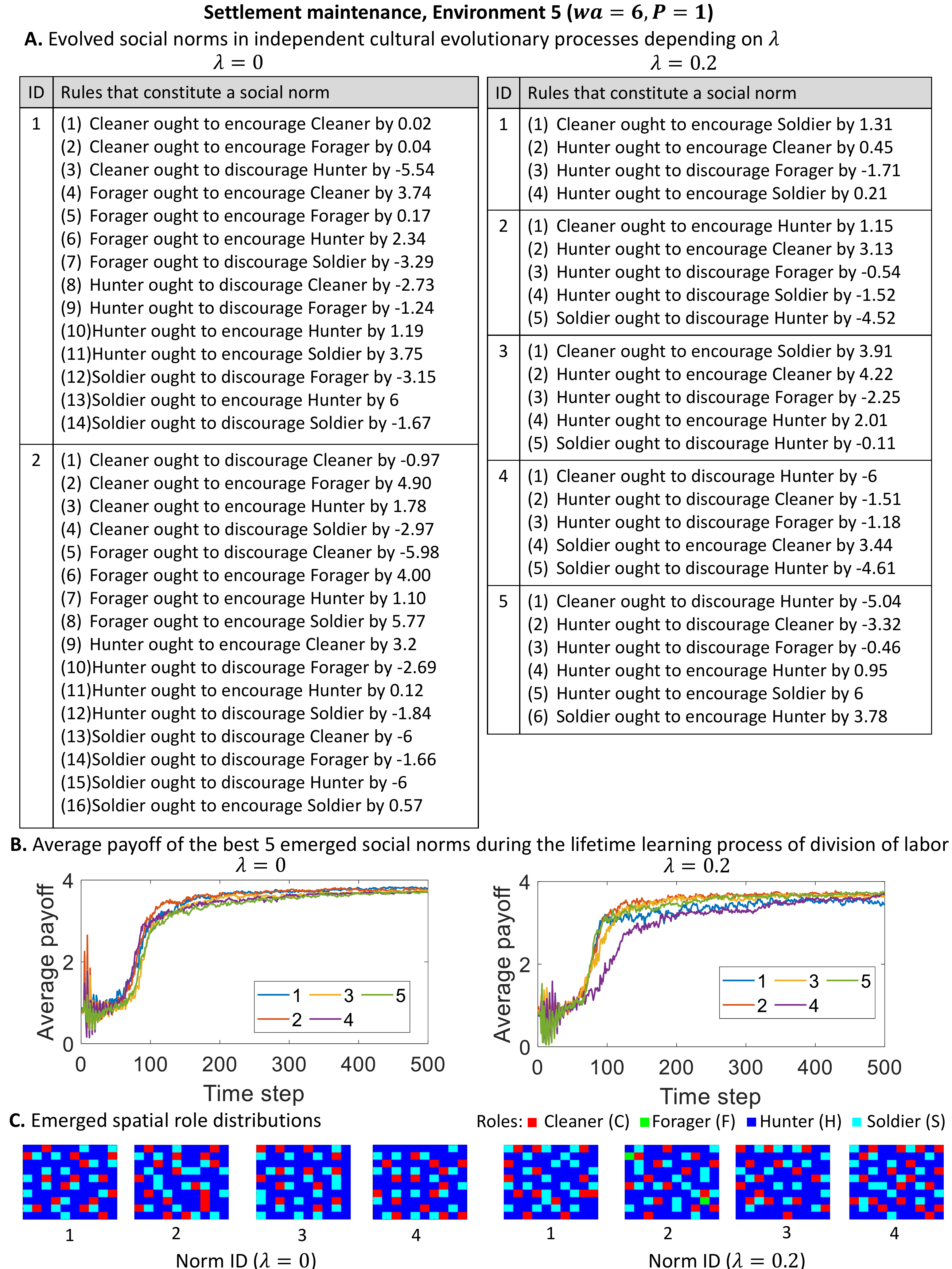}
\label{fig:Env5Settlement}
\end{figure*}

\begin{figure*}[!hb]
\centering
\includegraphics[width=0.95\textwidth]{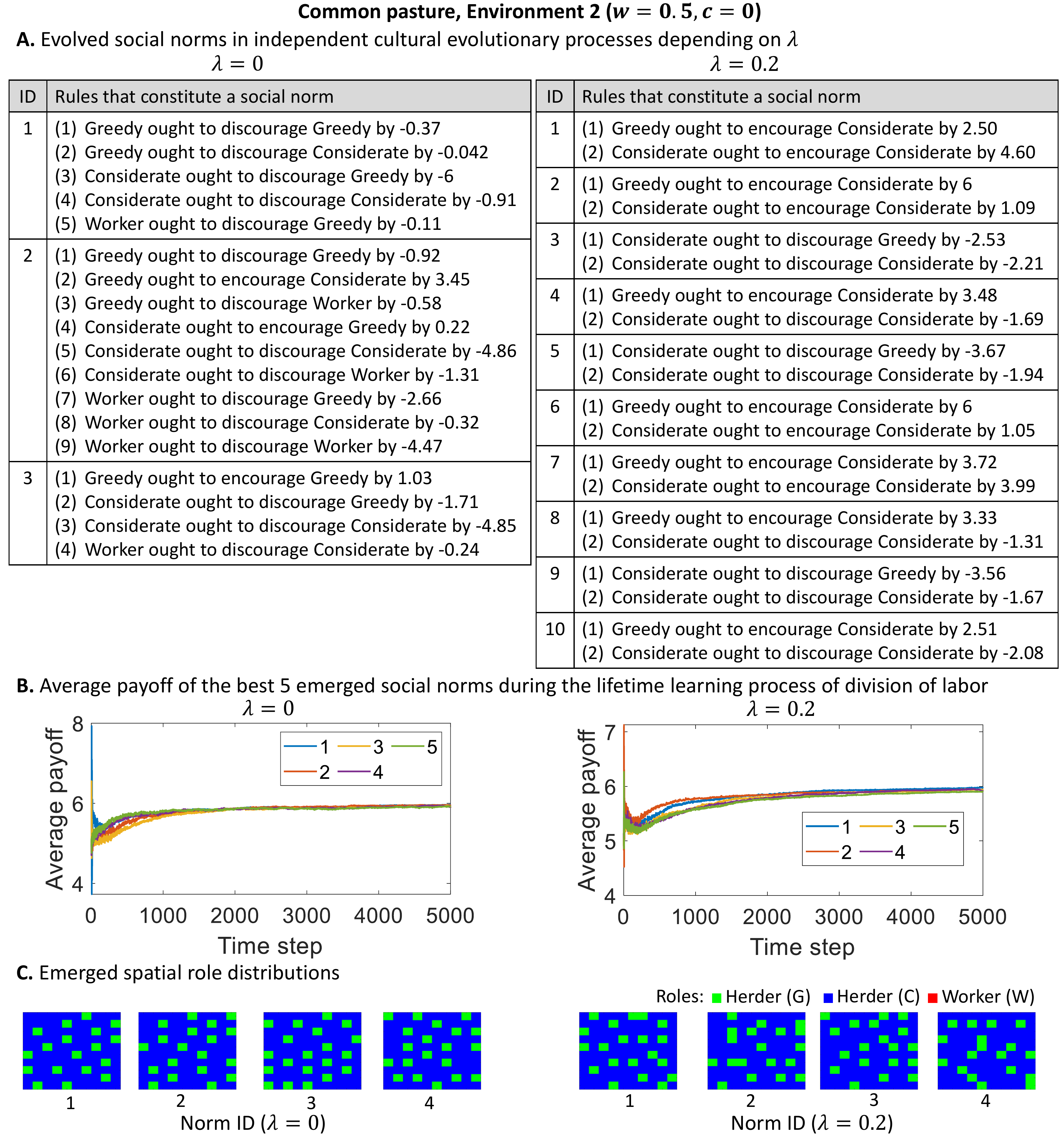}
\label{fig:Env2Pasture}
\end{figure*}

\begin{figure*}[!hb]
\centering
\includegraphics[width=0.95\textwidth]{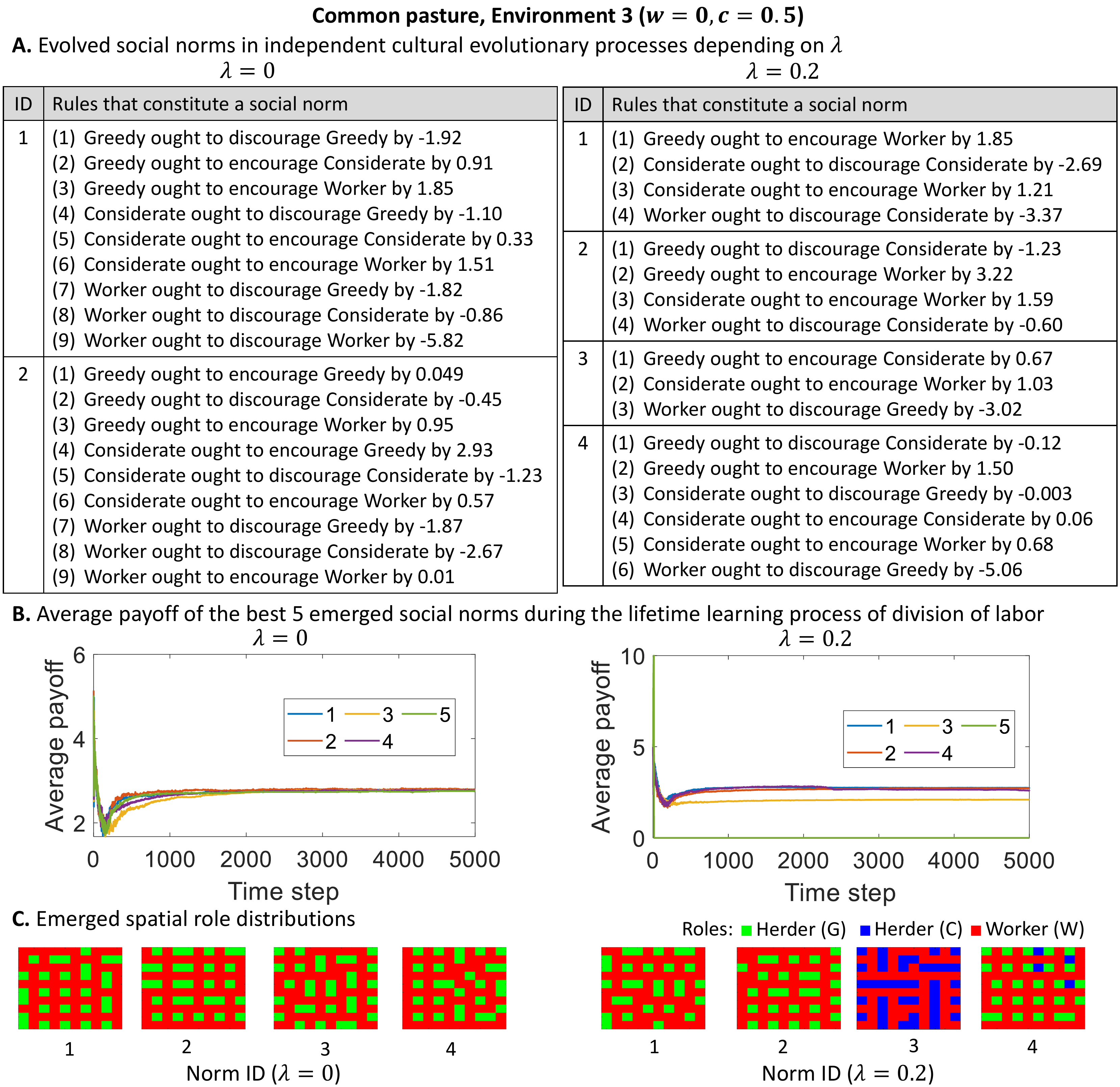}
\label{fig:Env3Pasture}
\end{figure*}

\begin{figure*}[!hb]
\centering
\includegraphics[width=0.95\textwidth]{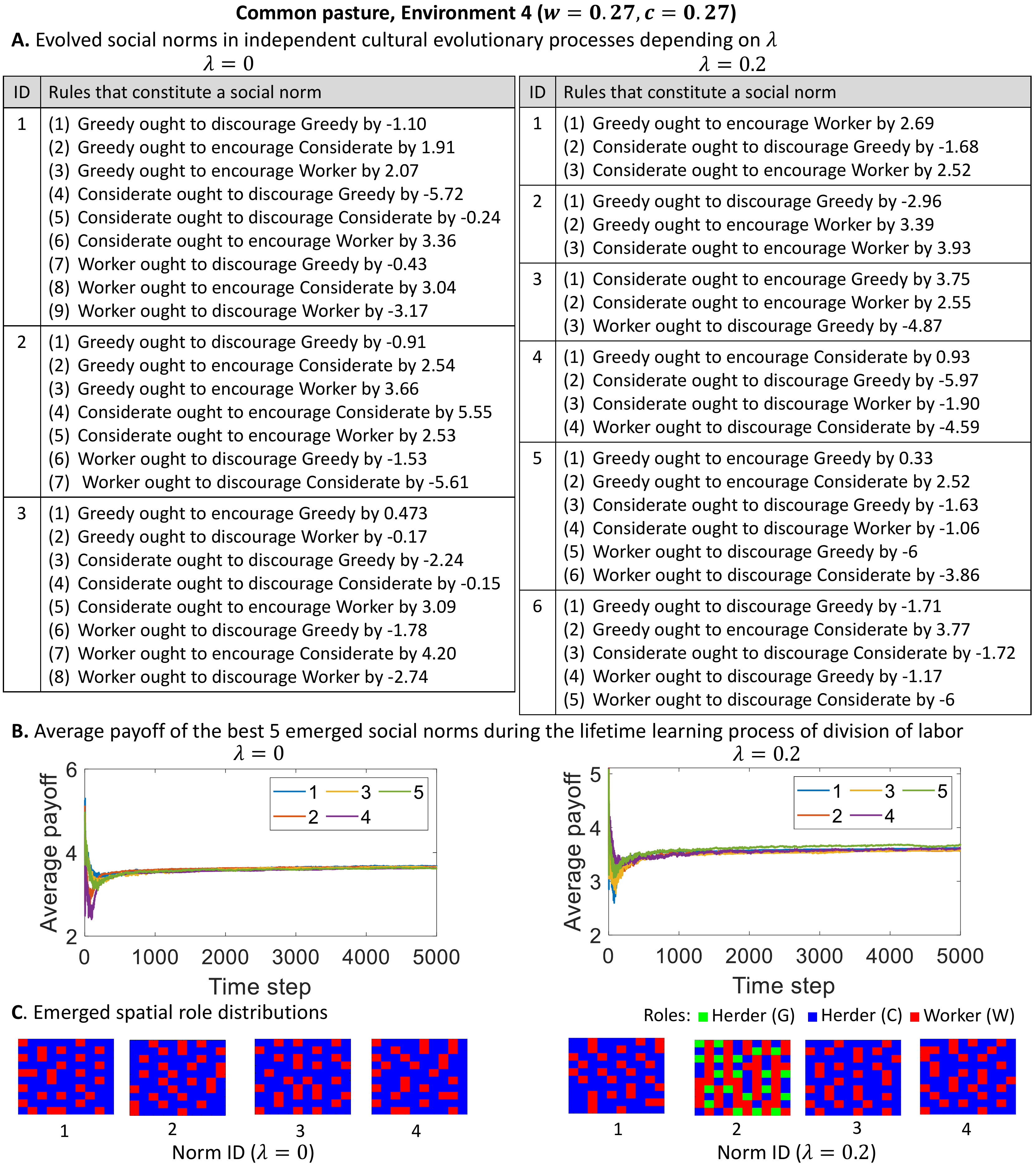}
\label{fig:Env4Pasture}
\end{figure*}

\begin{figure*}[!hb]
\centering
\includegraphics[width=0.95\textwidth]{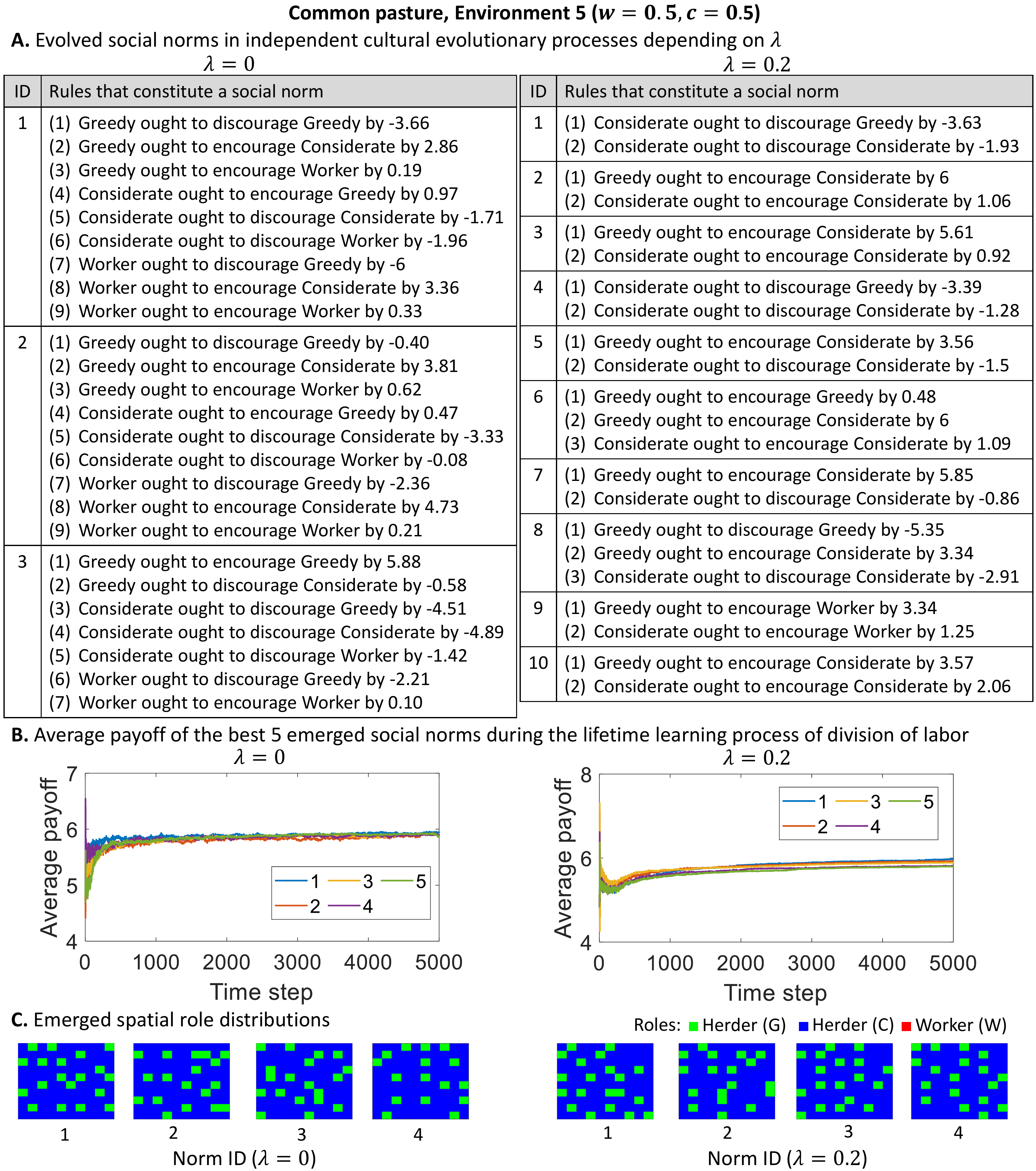}
\label{fig:Env5Pasture}
\end{figure*}

\end{document}